# A New Variable in General Relativity and Its Applications for Classic and Quantum Gravity

T. Mei

(Central China Normal University, Wuhan, Hubei PRO, People's Republic of China

E-Mail:   meitao@mail.ccnu.edu.cn     meitaowh@public.wh.hb.cn )

**Abstract:**   A new variable in the Riemannian geometry is introduced by the tetrad and the Ricci's coefficients of rotation, the characters of *curve* of the Riemannian geometry are determined completely by the new variable; for general relativity, all the Einstein-Hilbert action, the Einstein equation in general relativity and the Dirac equation in curved spacetime can be expressed by the new variable, and, further, as well the action of the theory on the interaction of gravitational, electromagnetic and spinor field (TGESF). All the characters of transformations of the new variable, the Einstein-Hilbert action and the action of TGESF under the general coordinate transformations and the local Lorentz transformations are discussed, respectively. After presenting the method of introduction of gravitational field in terms of the principle of gauge invariance based on the Dirac equation, and the ten constraint conditions for the tetrad are given, the vacuum-vacuum transition amplitude with the Faddeev-Popov ghost and the terms of external sources of the pure gravitational field is presented; finally, as well that of TGESF.

## 1   Introduction

As well known, there are two kinds of basic variables for expressing general relativity: the metric tensor $g_{\mu\nu}$ and the tetrad field $e_\mu^{\hat{\alpha}}$, which have the relations

$$g_{\mu\nu} = \eta_{\hat{\alpha}\hat{\beta}} e_\mu^{\hat{\alpha}} e_\nu^{\hat{\beta}} = e_\mu^{\hat{\alpha}} e_{\hat{\alpha}\nu} . \tag{1}$$

We use the Greek alphabet to denote spacetime indices, and the Greek alphabet with the symbol "^" to denote the local frame components, the Greek alphabet with the symbol "^" is raised or lowered by $\eta^{\hat{\alpha}\hat{\beta}} = \mathrm{diag}(-1,+1,+1,+1)$ or $\eta_{\hat{\alpha}\hat{\beta}}$, respectively. The basic characters of $e_\mu^{\hat{\alpha}}$ are represented here:

$$e_{\hat{\alpha}}^\mu = \eta_{\hat{\alpha}\hat{\beta}} g^{\mu\nu} e_\nu^{\hat{\beta}} , \; e_\mu^{\hat{\alpha}} e_{\hat{\alpha}}^\nu = \delta_\mu^\nu , \; e_\mu^{\hat{\beta}} e_{\hat{\alpha}}^\mu = \delta_{\hat{\alpha}}^{\hat{\beta}} . \tag{2}$$

On the other hand, there has been great and continual interest for seeking new variables to express the theory of gravity. For example, the canonical form of the Einstein-Hilbert action using the Ricci's coefficients of rotation[1], the loop quantum gravity based on the Ashtekar variables[2], the teleparallel gravity based on the Weitzenböck connection[3], the theory of spin-2 field using the Fierz tensor[4], and some variables in the gauge theory of gravity and supergravity[5, 6] and that of super-string, etc.

In this paper, we introduce a new variable $F_{\mu\nu\lambda}$ in the Riemannian geometry in terms of the tetrad and the Ricci's coefficients of rotation. The new variable $F_{\mu\nu\lambda}$ has the following main

characters:

① Its definition is very simple, in fact, $F_{\mu\nu\lambda}$ is only a re-representation of the Ricci's coefficients of rotation in the global manifold coordinate system.

② Its function is strong: As viewed from mathematics, the characters of *curve* of the Riemannian geometry are determined completely by $F_{\mu\nu\lambda}$, because the Riemann curvature tensor $R^{\mu}_{\rho\nu\sigma}$ can be expressed by $F_{\mu\nu\lambda}$; on the other hand, for physics, especially general relativity, all the Einstein-Hilbert action, the Einstein equation in general relativity and the Dirac equation in curved spacetime can be expressed by $F_{\mu\nu\lambda}$, and, further, as well the action of the theory on the interaction of gravitational, electromagnetic and spinor field (abbreviated TGESF).

③ Its behave is well, because some quantities expressed by $F_{\mu\nu\lambda}$ have well characters under coordinate transformations.

There are two kinds of coordinate transformations: The general coordinate transformations of the global manifold coordinate system when the local frame coordinate system is fixed (abbreviated T-1), and, contrarily, the local Lorentz transformations of the local frame coordinate system when the global manifold coordinate system is fixed (abbreviated T-2).

Just as the Yang-Mills field can be introduced by the principle of gauge invariance based on the Dirac equation, gravitational field whose Einstein-Hilbert action is expressed by $F_{\mu\nu\lambda}$ can be introduced by the principle of invariance under T-1 and T-2 based on the Dirac equation as well. And, further, all the characters of transformations under T-1 and T-2 of the Einstein-Hilbert action and the action of TGESF expressed by $F_{\mu\nu\lambda}$ are so well that we can use directly the technologies of the modern gauge theory of quantized field to realize quantization of the pure gravitational field and, further, to establish a quantized TGESF. The quantum theory of the pure gravitational field established in present paper is different from the earlier covariant quantization form for general relativity[7].

## 2　The expression for classic general relativity in terms of a new variable
### 2.1　Introduction of the new variable

At first, we represent some results in Ref. [8].

As is well-known the Einstein-Hilbert action

$$L_{\text{EH}} = \int R\sqrt{-g}\,\mathrm{d}^4 x = \int L_{g-\text{den}}\sqrt{-g}\,\mathrm{d}^4 x + \int \frac{\partial w^{\mu}_{(1)}(x)}{\partial x^{\mu}}\mathrm{d}^4 x, \tag{3}$$

$$L_{g-\text{den}} = g^{\alpha\beta}(\Gamma^{\rho}_{\alpha\sigma}\Gamma^{\sigma}_{\beta\rho} - \Gamma^{\rho}_{\alpha\beta}\Gamma^{\sigma}_{\rho\sigma}); \tag{4}$$

On the other hand, the Riemann curvature tensor $R^{\mu}_{\rho\nu\sigma} = \Gamma^{\mu}_{\rho\sigma,\nu} - \Gamma^{\mu}_{\rho\nu,\sigma} + \Gamma^{\mu}_{\lambda\nu}\Gamma^{\lambda}_{\rho\sigma} - \Gamma^{\mu}_{\lambda\sigma}\Gamma^{\lambda}_{\rho\nu}$ can be expressed by the tetrad field $e^{\hat{\alpha}}_{\mu}$:

$$R^{\mu}_{\rho\nu\sigma} = e^{\hat{\alpha}}_{\rho}(e^{\mu}_{\hat{\alpha};\sigma;\nu} - e^{\mu}_{\hat{\alpha};\nu;\sigma}), \tag{5}$$

where ";" means the covariant derivative, e.g., $e^{\mu}_{\hat{\alpha};\sigma} = e^{\mu}_{\hat{\alpha},\sigma} + e^{\lambda}_{\hat{\alpha}} \Gamma^{\mu}_{\lambda\sigma}$. And, using the formula $e^{\hat{\alpha}\mu}{}_{;\lambda} e^{\nu}_{\hat{\alpha}} = -e^{\hat{\alpha}\mu} e^{\nu}_{\hat{\alpha};\lambda}$ which can be derived from $g^{\mu\nu}{}_{;\lambda} = (e^{\hat{\alpha}\mu} e^{\nu}_{\hat{\alpha}})_{;\lambda} = 0$, we obtain

$$\sqrt{-g}R = \sqrt{-g}g^{\rho\sigma}R_{\rho\sigma} = \sqrt{-g}g^{\rho\sigma}R^{\mu}_{\rho\mu\sigma} = \sqrt{-g}e^{\hat{\alpha}\mu}(e^{\nu}_{\hat{\alpha};\mu;\nu} - e^{\nu}_{\hat{\alpha};\nu;\mu})$$
$$= \sqrt{-g}(e^{\hat{\alpha}\mu}{}_{;\mu} e^{\nu}_{\hat{\alpha};\nu} - e^{\hat{\alpha}\mu}{}_{;\nu} e^{\nu}_{\hat{\alpha};\mu}) - 2\frac{\partial}{\partial x^{\mu}}\left(\sqrt{-g}e^{\hat{\alpha}\mu}e^{\nu}_{\hat{\alpha};\nu}\right);$$

Substituting the above formula to (3), we have

$$L_{\text{EH}} = \int L_{\text{G-den}}\sqrt{-g}\,d^4x + \int \frac{\partial w^{\mu}_{(2)}}{\partial x^{\mu}}d^4x = \int L_{\text{G-den}}\left(\det e^{\hat{\alpha}}_{\mu}\right)d^4x + \int \frac{\partial w^{\mu}_{(2)}(x)}{\partial x^{\mu}}d^4x, \quad (6)$$

$$L_{\text{G-den}} = e^{\hat{\alpha}\mu}{}_{;\mu} e^{\nu}_{\hat{\alpha};\nu} - e^{\hat{\alpha}\mu}{}_{;\nu} e^{\nu}_{\hat{\alpha};\mu}. \quad (7)$$

Notice the difference between (7) and (4): $L_{\text{G-den}}$ is a scalar in the Riemannian geometry but $L_{g-\text{den}}$ to be not, the remainder of $L_{g-\text{den}}$ and $L_{\text{G-den}}$ is a total derivative:

$$L_{g-\text{den}} = L_{\text{G-den}} + \frac{\partial}{\partial x^{\mu}}\left[\sqrt{-g}\left(e^{\hat{\alpha}\mu}\frac{\partial e^{\nu}_{\hat{\alpha}}}{\partial x^{\nu}} - e^{\hat{\alpha}\nu}\frac{\partial e^{\mu}_{\hat{\alpha}}}{\partial x^{\nu}}\right)\right].$$

Using the Ricci's coefficients of rotation[8-11]

$$r_{\hat{\alpha}\hat{\beta}\hat{\gamma}} = e_{\hat{\alpha}\mu;\nu} e^{\mu}_{\hat{\beta}} e^{\nu}_{\hat{\gamma}} \quad (8)$$

(7) can be rewritten in the form:

$$L_{\text{G-den}} = r^{\hat{\alpha}\hat{\beta}}{}_{\hat{\beta}} r_{\hat{\alpha}}{}^{\hat{\gamma}}{}_{\hat{\gamma}} - r^{\hat{\alpha}\hat{\beta}\hat{\gamma}} r_{\hat{\alpha}\hat{\gamma}\hat{\beta}}, \quad (9)$$

The above contents (from (3) to (9)) can be found in Ref.[8], the proof that the Einstein equation can be obtained by $L_{\text{G-den}}$ given by (7) or (9) in terms of the corresponding Euler-Lagrange equation can be found in Ref.[8] as well.

Based on these results in Ref.[8], we define a quantity

$$\begin{aligned}F_{\mu\nu\lambda} &= r_{\hat{\alpha}\hat{\beta}\hat{\gamma}} e^{\hat{\alpha}}_{\mu} e^{\hat{\beta}}_{\nu} e^{\hat{\gamma}}_{\lambda} = e^{\hat{\alpha}}_{\mu} e_{\hat{\alpha}\nu;\lambda} \\ &= \frac{1}{2}\eta_{\hat{\alpha}\hat{\beta}} e^{\hat{\alpha}}_{\mu} e^{\hat{\beta}}_{\nu,\lambda} - \frac{1}{2}\eta_{\hat{\alpha}\hat{\beta}} e^{\hat{\alpha}}_{\nu} e^{\hat{\beta}}_{\mu,\lambda} \\ &\quad + \frac{1}{2}\eta_{\hat{\alpha}\hat{\beta}} e^{\hat{\alpha}}_{\nu} e^{\hat{\beta}}_{\lambda,\mu} - \frac{1}{2}\eta_{\hat{\alpha}\hat{\beta}} e^{\hat{\alpha}}_{\mu} e^{\hat{\beta}}_{\lambda,\nu} \\ &\quad + \frac{1}{2}\eta_{\hat{\alpha}\hat{\beta}} e^{\hat{\alpha}}_{\lambda} e^{\hat{\beta}}_{\nu,\mu} - \frac{1}{2}\eta_{\hat{\alpha}\hat{\beta}} e^{\hat{\alpha}}_{\lambda} e^{\hat{\beta}}_{\mu,\nu}.\end{aligned} \quad (10)$$

We see that $F_{\mu\nu\lambda}$ is only a re-representation of the Ricci's coefficients of rotation in the global manifold coordinate system, and according to the above definition of $F_{\mu\nu\lambda}$ we can prove that raising or lowering Greek alphabet in $F_{\mu\nu\lambda}$ is done by the metric tensor $g^{\mu\nu}$ or $g_{\mu\nu}$, respectively; e.g., $F_{\mu\lambda}{}^{\lambda} = g^{\rho\sigma}F_{\mu\rho\sigma}$, etc.

It can be proved that $L_{\text{G-den}}$ given by (9) can be rewrite by $F_{\mu\nu\lambda}$:

$$L_{\text{G-den}} = F^{\mu\nu}{}_{\nu} F_{\mu\lambda}{}^{\lambda} - F^{\mu\nu\lambda} F_{\mu\lambda\nu}, \quad (11)$$

and using the standard method deriving the Einstein equation in terms of the tetrad field $e_\mu^{\hat{\alpha}}$ [8, 12] we can prove that $L_{\text{G-den}}$ given by (11) leads to the Einstein equation.

The characters of *curve* of the Riemannian geometry is determined completely by the Ricci's coefficients of rotation[9, 10], in other words, by $F_{\mu\nu\lambda}$.

Analogously to the character that $r_{\hat{\alpha}\hat{\beta}\hat{\gamma}}$ is antisymmetric in the first pair of indices:

$$r_{\hat{\alpha}\hat{\beta}\hat{\gamma}} + r_{\hat{\beta}\hat{\alpha}\hat{\gamma}} = 0, \tag{12}$$

$F_{\mu\nu\lambda}$ has the same character as well:

$$F_{\mu\nu\lambda} + F_{\nu\mu\lambda} = 0, \tag{13}$$

thus, the both $r_{\hat{\alpha}\hat{\beta}\hat{\gamma}}$ and $F_{\mu\nu\lambda}$ have at most 24 independent components in the 4-dimensional Riemannian geometry, respectively.

As for the last pair of indices of $F_{\mu\nu\lambda}$, from (10) we have

$$\begin{aligned} F_{\lambda\mu\nu} + F_{\lambda\nu\mu} &= \eta_{\hat{\alpha}\hat{\beta}} e_\mu^{\hat{\alpha}} e_{\nu,\lambda}^{\hat{\beta}} + \eta_{\hat{\alpha}\hat{\beta}} e_\nu^{\hat{\alpha}} e_{\mu,\lambda}^{\hat{\beta}} - \eta_{\hat{\alpha}\hat{\beta}} e_\mu^{\hat{\alpha}} e_{\lambda,\nu}^{\hat{\beta}} - \eta_{\hat{\alpha}\hat{\beta}} e_\nu^{\hat{\alpha}} e_{\lambda,\mu}^{\hat{\beta}}, \\ F_{\lambda\mu\nu} - F_{\lambda\nu\mu} &= \eta_{\hat{\alpha}\hat{\beta}} e_\lambda^{\hat{\alpha}} e_{\mu,\nu}^{\hat{\beta}} - \eta_{\hat{\alpha}\hat{\beta}} e_\lambda^{\hat{\alpha}} e_{\nu,\mu}^{\hat{\beta}}. \end{aligned} \tag{14}$$

According to the definition of the Weitzenböck connection[3]

$$W^\lambda{}_{\mu\nu} = e_{\hat{\alpha}}^\lambda e_{\mu,\nu}^{\hat{\alpha}}$$

we can prove

$$F^\lambda{}_{\mu\nu} = W^\lambda{}_{\mu\nu} - \Gamma^\lambda_{\mu\nu}.$$

## 2.2 The characters of transformations $F_{\mu\nu\lambda}$ and the action constructed by $F_{\mu\nu\lambda}$

There are two systems of coordinate here: the global manifold coordinate system $\{x^\mu\}$ and the local frame coordinate system $\{\xi^{\hat{\alpha}}\}$, because $e_\mu^{\hat{\alpha}} = \frac{\partial \xi^{\hat{\alpha}}}{\partial x^\mu}$ and $e_{\hat{\alpha}}^\mu = \frac{\partial x^\mu}{\partial \xi^{\hat{\alpha}}}$ are dependent on not only $\{x^\mu\}$ but also $\{\xi^{\hat{\alpha}}\}$; and, accordingly, there are two kinds of coordinate transformations: T-1 represented by $x^\mu = x^\mu(\tilde{x}^\nu)$ and T-2 represented by $\mathrm{d}\xi^{\hat{\alpha}} = \Lambda_{\hat{\beta}}^{\hat{\alpha}}(x)\mathrm{d}\tilde{\xi}^{\hat{\beta}}$, respectively. At first, we discuss the case of T-1. (The both definitions of T-1 and T-2 can be found in the Section 1, ③ of the present paper.)

In the case of T-1, as well known, $e_\mu^{\hat{\alpha}}$ changes according to $e_\mu^{\hat{\alpha}} = \frac{\partial \tilde{x}^\nu}{\partial x^\mu} \tilde{e}_\nu^{\hat{\alpha}}$, that means that $e_\mu^{\hat{\alpha}}$ are as four covariant vectors; the Ricci's coefficients of rotation $r_{\hat{\alpha}\hat{\beta}\hat{\gamma}}$ are scalars (There are 64 scalar functions, although at most 24 independent scalars in the 4-dimensional Riemannian geometry); and, further, $F_{\mu\nu\lambda}$ defined by (10) changes according to

$$F_{\mu\nu\lambda} = \frac{\partial \tilde{x}^\alpha}{\partial x^\mu} \frac{\partial \tilde{x}^\beta}{\partial x^\nu} \frac{\partial \tilde{x}^\gamma}{\partial x^\lambda} \tilde{F}_{\alpha\beta\gamma},$$

that means that $F_{\mu\nu\lambda}$ is a three-index covariant tensor in the global manifold coordinate system (when the local frame coordinate system is fixed).

Now we discuss the case of T-2. In this case, $e_\mu^{\hat{\alpha}}$ changes according to

$$e_\mu^{\hat{\alpha}} = \Lambda_{\hat{\beta}}^{\hat{\alpha}}(x) \tilde{e}_\mu^{\hat{\beta}}, \quad e_{\hat{\alpha}}^\mu = \overline{\Lambda}_{\hat{\alpha}}^{\hat{\beta}}(x) \tilde{e}_{\hat{\beta}}^\mu, \tag{15}$$

where $\Lambda_{\hat{\beta}}^{\hat{\alpha}} \overline{\Lambda}_{\hat{\alpha}}^{\hat{\gamma}} = \delta_{\hat{\beta}}^{\hat{\gamma}}$, $\Lambda_{\hat{\alpha}}^{\hat{\gamma}} \overline{\Lambda}_{\hat{\beta}}^{\hat{\alpha}} = \delta_{\hat{\beta}}^{\hat{\gamma}}$, $\Lambda_{\hat{\beta}}^{\hat{\alpha}} = \Lambda_{\hat{\beta}}^{\hat{\alpha}}(x)$ is a function of the global coordinate $\{x^\mu\}$, that means that $\Lambda_{\hat{\beta}}^{\hat{\alpha}}$ at different spacetime point maybe different, this is so called *the local Lorentz transformations*.

We can obtain the rule of the transformation of $F_{\mu\nu\lambda}$ from the (10) and (15):

$$F_{\mu\nu\lambda} = \tilde{F}_{\mu\nu\lambda} + \Lambda_{\hat{\beta}}^{\hat{\alpha}} \overline{\Lambda}_{\hat{\alpha},\lambda}^{\hat{\gamma}} \tilde{e}_\mu^{\hat{\beta}} \tilde{e}_{\hat{\gamma}\nu},$$
$$\tilde{F}_{\mu\nu\lambda} = \tilde{e}_\mu^{\hat{\alpha}} \tilde{e}_{\hat{\alpha}\nu;\lambda}. \tag{16}$$

We see that $F_{\mu\nu\lambda}$ is changed under T-2, of course, this is a foreseeable result.

On the other hand, in a general way, a theory is asked to be variable neither T-1 nor T-2, we therefore investigate that what forms of the action constructed by the quadratic terms of $F_{\mu\nu\lambda}$ can satisfies the conditions of invariability under T-1 and T-2.

At first, if the Lagrange density of a theory is a scalar in the global manifold coordinate system, then it is invariable under T-1. Because $F_{\mu\nu\lambda}$ is a three-index covariant tensor for the global manifold coordinate system when the local frame coordinate system is fixed, the most general form of scalar constructed by the quadratic terms of $F_{\mu\nu\lambda}$ satisfying $F_{\mu\nu\lambda} = -F_{\nu\mu\lambda}$ is:

$$L_{\text{den}}(F_{\mu\nu\lambda}) = \kappa + a F^{\mu\nu}{}_\nu F_{\mu\lambda}{}^\lambda + b F^{\mu\nu\lambda} F_{\mu\lambda\nu} + \varepsilon F^{\mu\nu\lambda} F_{\mu\nu\lambda}. \tag{17}$$

where $\kappa$ is a constant.

Under T-2, from (16) we know the action constructed by $L_{\text{den}}(F_{\mu\nu\lambda})$ changes according to:

$$L = \int L_{\text{den}}(F_{\mu\nu\lambda}) \sqrt{-g} \, d^4x = \int L_{\text{den}}(\tilde{F}_{\mu\nu\lambda}) \sqrt{-g} \, d^4x + \int \Delta \, d^4x, \tag{18}$$

$$\Delta = a\Delta_a + b\Delta_b + \varepsilon\Delta_\varepsilon;$$
$$\Delta_a = \sqrt{-g} \, (2 \Lambda_{\hat{\beta}}^{\hat{\alpha}} \overline{\Lambda}_{\hat{\alpha},\sigma}^{\hat{\gamma}} \tilde{e}_\mu^{\hat{\beta}} \tilde{e}_{\hat{\gamma}}^\sigma \tilde{F}^{\mu\rho}{}_\rho + \Lambda_{\hat{\beta}}^{\hat{\alpha}} \overline{\Lambda}_{\hat{\alpha},\rho}^{\hat{\gamma}} \tilde{e}^{\hat{\beta}\mu} \tilde{e}_{\hat{\gamma}}^\rho \Lambda_{\hat{\chi}}^{\hat{\theta}} \overline{\Lambda}_{\hat{\theta},\sigma}^{\hat{\tau}} \tilde{e}_\mu^{\hat{\chi}} \tilde{e}_{\hat{\tau}}^\sigma),$$
$$\Delta_b = \sqrt{-g} \, (2 \Lambda_{\hat{\beta}}^{\hat{\alpha}} \overline{\Lambda}_{\hat{\alpha},\nu}^{\hat{\gamma}} \tilde{e}_\mu^{\hat{\beta}} \tilde{e}_{\hat{\gamma}\lambda} \tilde{F}^{\mu\nu\lambda} + \Lambda_{\hat{\beta}}^{\hat{\alpha}} \overline{\Lambda}_{\hat{\alpha}}^{\hat{\gamma},\lambda} \tilde{e}^{\hat{\beta}\mu} \tilde{e}_{\hat{\gamma}}^\nu \Lambda_{\hat{\chi}}^{\hat{\theta}} \overline{\Lambda}_{\hat{\theta},\nu}^{\hat{\tau}} \tilde{e}_\mu^{\hat{\chi}} \tilde{e}_{\hat{\tau}\lambda}), \tag{19}$$
$$\Delta_\varepsilon = \sqrt{-g} \, (2 \Lambda_{\hat{\beta}}^{\hat{\alpha}} \overline{\Lambda}_{\hat{\alpha},\lambda}^{\hat{\gamma}} \tilde{e}_\mu^{\hat{\beta}} \tilde{e}_{\hat{\gamma}\nu} \tilde{F}^{\mu\nu\lambda} + \Lambda_{\hat{\beta}}^{\hat{\alpha}} \overline{\Lambda}_{\hat{\alpha}}^{\hat{\gamma},\lambda} \tilde{e}^{\hat{\beta}\mu} \tilde{e}_{\hat{\gamma}}^\nu \Lambda_{\hat{\chi}}^{\hat{\theta}} \overline{\Lambda}_{\hat{\theta},\lambda}^{\hat{\tau}} \tilde{e}_\mu^{\hat{\chi}} \tilde{e}_{\hat{\tau}\nu}).$$

For simplifying (19) we need the formulas on the Lorentz transformations:

$$\eta^{\hat{\alpha}\hat{\gamma}} \Lambda_{\hat{\gamma}}^{\hat{\beta}} = \eta^{\hat{\beta}\hat{\gamma}} \overline{\Lambda}_{\hat{\gamma}}^{\hat{\alpha}}, \quad \eta_{\hat{\alpha}\hat{\gamma}} \Lambda_{\hat{\beta}}^{\hat{\gamma}} = \eta_{\hat{\beta}\hat{\gamma}} \Lambda_{\hat{\alpha}}^{\hat{\gamma}}, \tag{20}$$

which can be obtained by considering

$$e_{\hat{\alpha}}^{\mu} = \eta_{\hat{\alpha}\hat{\gamma}} e^{\hat{\gamma}\mu} = \eta_{\hat{\alpha}\hat{\gamma}} \Lambda_{\hat{\chi}}^{\hat{\gamma}} \tilde{e}^{\hat{\chi}\mu} = \eta_{\hat{\alpha}\hat{\gamma}} \Lambda_{\hat{\chi}}^{\hat{\gamma}} \eta^{\hat{\chi}\hat{\tau}} \tilde{e}_{\hat{\tau}}^{\mu},$$

and comparing the above expression with (15).

Substituting the expression of $\widetilde{F}_{\mu\nu\lambda}$ given by (16) to (19), using (20) and $\tilde{e}^{\hat{\beta}\mu} \tilde{e}_{\mu}^{\hat{\chi}} = \eta^{\hat{\beta}\hat{\chi}}$ we can simplify $\Delta_a$, $\Delta_b$ and $\Delta_\varepsilon$. At first, for $\Delta_a$ we have:

$$\Delta_a = 2\sqrt{-g}\,\Lambda_{\hat{\beta}}^{\hat{\alpha}} \overline{\Lambda}_{\hat{\alpha},\rho}^{\hat{\gamma}} \tilde{e}_{\hat{\gamma}}^{\rho} \tilde{e}^{\hat{\beta}\sigma}{}_{,\sigma} + 2\frac{\partial \sqrt{-g}}{\partial x^{\sigma}} \Lambda_{\hat{\beta}}^{\hat{\alpha}} \overline{\Lambda}_{\hat{\alpha},\rho}^{\hat{\gamma}} \tilde{e}_{\hat{\gamma}}^{\rho} \tilde{e}^{\hat{\beta}\sigma} + \sqrt{-g}\,\eta^{\hat{\beta}\hat{\chi}} \Lambda_{\hat{\beta}}^{\hat{\alpha}} \Lambda_{\hat{\chi}}^{\hat{\theta}} \overline{\Lambda}_{\hat{\alpha},\rho}^{\hat{\gamma}} \overline{\Lambda}_{\hat{\theta},\sigma}^{\hat{\tau}} \tilde{e}_{\hat{\gamma}}^{\rho} \tilde{e}_{\hat{\tau}}^{\sigma}$$

$$= 2\frac{\partial}{\partial x^{\sigma}}\left(\sqrt{-g}\,\Lambda_{\hat{\beta}}^{\hat{\alpha}} \overline{\Lambda}_{\hat{\alpha},\rho}^{\hat{\gamma}} \tilde{e}_{\hat{\gamma}}^{\rho} \tilde{e}^{\hat{\beta}\sigma}\right) - 2\sqrt{-g}\,\Lambda_{\hat{\beta}}^{\hat{\alpha}} \overline{\Lambda}_{\hat{\alpha},\rho}^{\hat{\gamma}} \tilde{e}_{\hat{\gamma},\sigma}^{\rho} \tilde{e}^{\hat{\beta}\sigma} + \sqrt{-g}\,\Lambda_{\hat{\beta},\rho}^{\hat{\alpha}} \overline{\Lambda}_{\hat{\alpha},\sigma}^{\hat{\gamma}} \tilde{e}_{\hat{\gamma}}^{\rho} \tilde{e}^{\hat{\beta}\sigma} - \Delta_{a1},$$

$$\Delta_{a1} = 2\sqrt{-g}\,\Lambda_{\hat{\beta}}^{\hat{\alpha}} \overline{\Lambda}_{\hat{\alpha},\rho,\sigma}^{\hat{\gamma}} \tilde{e}_{\hat{\gamma}}^{\rho} \tilde{e}^{\hat{\beta}\sigma} + \sqrt{-g}\,\Lambda_{\hat{\beta},\sigma}^{\hat{\alpha}} \overline{\Lambda}_{\hat{\alpha},\rho}^{\hat{\gamma}} \tilde{e}_{\hat{\gamma}}^{\rho} \tilde{e}^{\hat{\beta}\sigma} + \sqrt{-g}\,\Lambda_{\hat{\beta},\rho}^{\hat{\alpha}} \overline{\Lambda}_{\hat{\alpha},\sigma}^{\hat{\gamma}} \tilde{e}_{\hat{\gamma}}^{\rho} \tilde{e}^{\hat{\beta}\sigma},$$

using (20) we have:

$$\Delta_{a1} = \sqrt{-g}\,(\Lambda_{\hat{\beta}}^{\hat{\alpha}} \overline{\Lambda}_{\hat{\alpha}}^{\hat{\gamma}})_{,\rho,\sigma} \tilde{e}_{\hat{\gamma}}^{\rho} \tilde{e}^{\hat{\beta}\sigma} = \sqrt{-g}\,(\delta_{\hat{\beta}}^{\hat{\gamma}})_{,\rho,\sigma} \tilde{e}_{\hat{\gamma}}^{\rho} \tilde{e}^{\hat{\beta}\sigma} = 0.$$

For $\Delta_b$, notice $\Lambda_{\hat{\beta}}^{\hat{\alpha}} \overline{\Lambda}_{\hat{\alpha},\nu}^{\hat{\gamma}} = -\Lambda_{\hat{\beta},\nu}^{\hat{\alpha}} \overline{\Lambda}_{\hat{\alpha}}^{\hat{\gamma}}$ which can be obtained by $(\Lambda_{\hat{\beta}}^{\hat{\alpha}} \overline{\Lambda}_{\hat{\alpha}}^{\hat{\gamma}})_{,\nu} = (\delta_{\hat{\beta}}^{\hat{\gamma}})_{,\nu} = 0$ we have

$$\Delta_b = -2\sqrt{-g}\,\Lambda_{\hat{\beta}}^{\hat{\alpha}} \overline{\Lambda}_{\hat{\alpha},\rho}^{\hat{\gamma}} \tilde{e}_{\hat{\gamma},\sigma}^{\rho} \tilde{e}^{\hat{\beta}\sigma} + 2\sqrt{-g}\,\Lambda_{\hat{\beta}}^{\hat{\alpha}} \overline{\Lambda}_{\hat{\alpha},\rho}^{\hat{\gamma}} \tilde{e}_{\hat{\gamma}}^{\sigma} \tilde{e}^{\hat{\beta}\lambda} \Gamma_{\lambda\sigma}^{\rho} + \sqrt{-g}\,\Lambda_{\hat{\beta},\rho}^{\hat{\alpha}} \overline{\Lambda}_{\hat{\alpha},\sigma}^{\hat{\gamma}} \tilde{e}_{\hat{\gamma}}^{\rho} \tilde{e}^{\hat{\beta}\sigma},$$

As for the middle term,

$$\Lambda_{\hat{\beta}}^{\hat{\alpha}} \overline{\Lambda}_{\hat{\alpha},\rho}^{\hat{\gamma}} \tilde{e}_{\hat{\gamma}}^{\sigma} \tilde{e}^{\hat{\beta}\lambda} \Gamma_{\lambda\sigma}^{\rho} = \eta^{\hat{\beta}\hat{\chi}} \Lambda_{\hat{\beta}}^{\hat{\alpha}} \eta_{\hat{\gamma}\hat{\tau}} \overline{\Lambda}_{\hat{\alpha},\rho}^{\hat{\gamma}} \tilde{e}^{\hat{\tau}\sigma} \tilde{e}_{\hat{\chi}}^{\lambda} \Gamma_{\lambda\sigma}^{\rho} = \eta^{\hat{\alpha}\hat{\beta}} \overline{\Lambda}_{\hat{\beta}}^{\hat{\chi}} \eta_{\hat{\alpha}\hat{\gamma}} \Lambda_{\hat{\tau},\rho}^{\hat{\gamma}} \tilde{e}^{\hat{\tau}\sigma} \tilde{e}_{\hat{\chi}}^{\lambda} \Gamma_{\lambda\sigma}^{\rho}$$

$$= \Lambda_{\hat{\beta},\rho}^{\hat{\alpha}} \overline{\Lambda}_{\hat{\alpha}}^{\hat{\gamma}} \tilde{e}^{\hat{\beta}\sigma} \tilde{e}_{\hat{\gamma}}^{\lambda} \Gamma_{\lambda\sigma}^{\rho} = \Lambda_{\hat{\beta},\rho}^{\hat{\alpha}} \overline{\Lambda}_{\hat{\alpha}}^{\hat{\gamma}} \tilde{e}^{\hat{\beta}\lambda} \tilde{e}_{\hat{\gamma}}^{\sigma} \Gamma_{\sigma\lambda}^{\rho},$$

we therefore have

$$2\sqrt{-g}\,\Lambda_{\hat{\beta}}^{\hat{\alpha}} \overline{\Lambda}_{\hat{\alpha},\rho}^{\hat{\gamma}} \tilde{e}_{\hat{\gamma}}^{\sigma} \tilde{e}^{\hat{\beta}\lambda} \Gamma_{\lambda\sigma}^{\rho} = \sqrt{-g}\,\Lambda_{\hat{\beta}}^{\hat{\alpha}} \overline{\Lambda}_{\hat{\alpha},\rho}^{\hat{\gamma}} \tilde{e}_{\hat{\gamma}}^{\sigma} \tilde{e}^{\hat{\beta}\lambda} \Gamma_{\lambda\sigma}^{\rho} + \sqrt{-g}\,\Lambda_{\hat{\beta},\rho}^{\hat{\alpha}} \overline{\Lambda}_{\hat{\alpha}}^{\hat{\gamma}} \tilde{e}^{\hat{\beta}\lambda} \tilde{e}_{\hat{\gamma}}^{\sigma} \Gamma_{\sigma\lambda}^{\rho}$$

$$= \sqrt{-g}\,(\Lambda_{\hat{\beta}}^{\hat{\alpha}} \overline{\Lambda}_{\hat{\alpha},\rho}^{\hat{\gamma}} + \Lambda_{\hat{\beta},\rho}^{\hat{\alpha}} \overline{\Lambda}_{\hat{\alpha}}^{\hat{\gamma}}) \tilde{e}_{\hat{\gamma}}^{\sigma} \tilde{e}^{\hat{\beta}\lambda} \Gamma_{\lambda\sigma}^{\rho} = \sqrt{-g}\,(\Lambda_{\hat{\beta}}^{\hat{\alpha}} \overline{\Lambda}_{\hat{\alpha}}^{\hat{\gamma}})_{,\rho} \tilde{e}_{\hat{\gamma}}^{\sigma} \tilde{e}^{\hat{\beta}\lambda} \Gamma_{\lambda\sigma}^{\rho}$$

$$= \sqrt{-g}\,(\delta_{\hat{\beta}}^{\hat{\gamma}})_{,\rho} \tilde{e}_{\hat{\gamma}}^{\sigma} \tilde{e}^{\hat{\beta}\lambda} \Gamma_{\lambda\sigma}^{\rho} = 0.$$

For $\Delta_\varepsilon$ we have

$$\Delta_\varepsilon = 2\sqrt{-g}\,\Lambda_{\hat{\beta}}^{\hat{\alpha}} \overline{\Lambda}_{\hat{\alpha}}^{\hat{\gamma},\rho} \tilde{e}^{\hat{\beta}\sigma}{}_{,\rho} \tilde{e}_{\hat{\gamma}\sigma} + 2\sqrt{-g}\,\Lambda_{\hat{\beta}}^{\hat{\alpha}} \overline{\Lambda}_{\hat{\alpha}}^{\hat{\gamma},\rho} \tilde{e}^{\hat{\beta}\lambda} \tilde{e}_{\hat{\gamma}\sigma} \Gamma_{\lambda\rho}^{\sigma} + \sqrt{-g}\,\Lambda_{\hat{\beta},\lambda}^{\hat{\alpha}} \overline{\Lambda}_{\hat{\alpha}}^{\hat{\beta},\lambda}$$

$$= \sqrt{-g}\,\{2[\Lambda_{\hat{\beta}}^{\hat{\alpha}} \overline{\Lambda}_{\hat{\alpha},\rho}^{\hat{\gamma}} \tilde{e}^{\hat{\beta}\rho}{}_{,\sigma} \tilde{e}_{\hat{\gamma}}^{\sigma} - \Lambda_{\hat{\beta}}^{\hat{\alpha}} \overline{\Lambda}_{\hat{\alpha}}^{\hat{\gamma},\rho}(\tilde{e}_{\sigma}^{\hat{\beta}} \tilde{e}_{\hat{\gamma},\rho}^{\sigma} + \tilde{e}_{\hat{\gamma}}^{\sigma} \tilde{e}_{\rho,\sigma}^{\hat{\beta}})] + \Lambda_{\hat{\beta},\lambda}^{\hat{\alpha}} \overline{\Lambda}_{\hat{\alpha}}^{\hat{\beta},\lambda}\} \quad (21)$$

Summarizing the above results, we obtain the term $\int \Delta d^4 x$ in (18):

$$\int \Delta d^4 x = 2a \int \frac{\partial W^{\mu}(x)}{\partial x^{\mu}} d^4 x$$

$$+ (a+b) \int \sqrt{-g}\,(\Lambda_{\hat{\beta},\rho}^{\hat{\alpha}} \overline{\Lambda}_{\hat{\alpha},\sigma}^{\hat{\gamma}} \tilde{e}_{\hat{\gamma}}^{\rho} \tilde{e}^{\hat{\beta}\sigma} - 2\Lambda_{\hat{\beta}}^{\hat{\alpha}} \overline{\Lambda}_{\hat{\alpha},\rho}^{\hat{\gamma}} \tilde{e}_{\hat{\gamma},\sigma}^{\rho} \tilde{e}^{\hat{\beta}\sigma}) d^4 x + \varepsilon \int \Delta_\varepsilon d^4 x, \quad (22)$$

$$W^{\mu}(x) = \sqrt{-g}\,\Lambda_{\hat{\beta}}^{\hat{\alpha}} \overline{\Lambda}_{\hat{\alpha},\lambda}^{\hat{\gamma}} \tilde{e}_{\hat{\gamma}}^{\lambda} \tilde{e}^{\hat{\beta}\mu}.$$

We see, if $a+b=0, \varepsilon=0$, then change of the action $L$ given by (18) that is constructed by the Lagrange density $L_{\text{den}}(F_{\mu\nu\lambda})$ given by (17) under T-2 is just only an integral of a total derivative. However, $L_{\text{den}}(F_{\mu\nu\lambda})$ is just (11) with the cosmological constant $\kappa$ under the condition $a+b=0, \varepsilon=0$. In this paper we don't take account of the cosmological constant, if we must take account of it, what we have to do is just only adding $\kappa$ into (11). We therefore have proved that the change of $L_{\text{G-den}}$ given by (11) under T-2 is only to raise an integral of a total derivative in Einstein-Hilbert action, which does not impact on the derivation of the equation of motion.

A question raised from the above discussion is whether there exists a quantity $\Omega$ by which a scalar $L_{\text{den}}(\Omega)$ can be constructed in the global manifold coordinate system, and, further, $L_{\text{den}}(\Omega)$ changes according to

$$L_{\text{den}}(\Omega) \to L_{\text{den}}(\Phi) - \Delta_\varepsilon + \frac{\partial w^\mu_{(3)}(x)}{\partial x^\mu}$$

under T-2, where $\Delta_\varepsilon$ is given by (21), and $w^\mu_{(3)}(x)$ is an arbitrary function. We don't ask such $\Omega$ to be constructed by certain quantities in the general relativity; it might be an *extra field*. If such $\Omega$ could be found out, then the term $\varepsilon F^{\mu\nu\lambda} F_{\mu\nu\lambda} + \varepsilon L_{\text{den}}(\Omega)$ could be added into (11).

We can compare the characters of $L_{g-\text{den}}$ given by (4) with that of $L_{\text{G-den}}$ given by (11):

① Either $g_{\mu\nu}$ or $e^{\hat{\alpha}}_\mu$ can expresses $L_{g-\text{den}}$, but only $e^{\hat{\alpha}}_\mu$ can expresses completely $L_{\text{G-den}}$.

② There are $e^{\hat{\alpha}}_\mu$ and the first derivative of $e^{\hat{\alpha}}_\mu$, but not the second derivative of $e^{\hat{\alpha}}_\mu$, in the both expressions of $L_{g-\text{den}}$ and $L_{\text{G-den}}$, similarly, there are $g_{\mu\nu}$ and the first derivative of $g_{\mu\nu}$, but not the second derivative of $g_{\mu\nu}$, in the expression of $L_{g-\text{den}}$ expressed by $g_{\mu\nu}$.

③ $L_{g-\text{den}}$ is invariable under T-2 but variable under T-1, contrarily, $L_{\text{G-den}}$ is invariable under T-1 but variable under T-2. The both changes are only to raise an integral of a total derivative which is dynamically irrelevant in Einstein-Hilbert action, respectively.

Another question raised from the above discussion is whether one can find out a quantity $\Theta$ which has the following characters: (i) the Lagrange density $L_{\text{den}}(\Theta)$ constructed by $\Theta$ can recur perfectly general relativity, especially, by which the Einstein equation can be obtained; (ii) $L_{\text{den}}(\Theta)$ is variable neither T-1 nor T-2.

### 2.3 The expressions of the Riemannian curvature tensor and the Einstein equation

For the Riemann curvature tensor $R^\mu_{\rho\nu\sigma}$, according to (5) and (10) we have

$$R^{\mu}_{\rho\nu\sigma} = \left(e^{\hat{\alpha}}_{\rho} e^{\mu}_{\hat{\alpha};\sigma}\right)_{;\nu} - \left(e^{\hat{\alpha}}_{\rho} e^{\mu}_{\hat{\alpha};\nu}\right)_{;\sigma} - e^{\hat{\alpha}}_{\rho;\nu} e^{\mu}_{\hat{\alpha};\sigma} + e^{\hat{\alpha}}_{\rho;\sigma} e^{\mu}_{\hat{\alpha};\nu}$$

$$= F_{\rho}{}^{\mu}{}_{\sigma;\nu} - F_{\rho}{}^{\mu}{}_{\nu;\sigma} + F_{\lambda\rho\sigma} F^{\lambda\mu}{}_{\nu} - F_{\lambda\rho\nu} F^{\lambda\mu}{}_{\sigma},$$

$$R_{\mu\rho\nu\sigma} = g_{\mu\lambda} R^{\lambda}_{\rho\nu\sigma} = F_{\rho\mu\sigma;\nu} - F_{\rho\mu\nu;\sigma} + F_{\lambda\rho\sigma} F^{\lambda}{}_{\mu\nu} - F_{\lambda\rho\nu} F^{\lambda}{}_{\mu\sigma},$$

$$R_{\mu\nu} = R^{\lambda}_{\mu\lambda\nu} = F_{\mu}{}^{\lambda}{}_{\nu;\lambda} - F_{\mu}{}^{\lambda}{}_{\lambda;\nu} + F_{\rho\mu\nu} F^{\rho\sigma}{}_{\sigma} - F_{\rho\mu\sigma} F^{\rho\sigma}{}_{\nu},$$

$$R^{\mu\nu} = F^{\mu\lambda\nu}{}_{;\lambda} - F^{\mu\lambda}{}_{\lambda}{}^{;\nu} + F^{\rho\mu\nu} F_{\rho\sigma}{}^{\sigma} - F^{\rho\mu\sigma} F_{\rho\sigma}{}^{\nu},$$

$$R = g_{\mu\nu} R^{\mu\nu} = F^{\rho\sigma}{}_{\rho;\sigma} - F^{\rho\sigma}{}_{\sigma;\rho} + F^{\rho\sigma}{}_{\sigma} F_{\rho\lambda}{}^{\lambda} - F^{\rho\sigma\lambda} F_{\rho\lambda\sigma}.$$

(23)

Because $F_{\mu\nu\lambda}$ is a tensor in the global manifold coordinate system (when the local frame coordinate system is fixed), the operation of the covariant derivative for $F_{\mu\nu\lambda}$ is significative.

We see that the Riemann curvature tensor $R^{\mu}_{\rho\nu\sigma}$ is determined completely by $F_{\mu\nu\lambda}$, that means that the characters of *curve* of the Riemannian geometry are determined completely by $F_{\mu\nu\lambda}$.

As well known, $R$ is a scalar not only in the global manifold coordinate system, but also in the local frame coordinate system. On the other hand, notice $F^{\mu\nu\lambda} = -F^{\nu\mu\lambda}$, we can rewrite $\sqrt{-g} R$ in the form:

$$\sqrt{-g} R = -2 \frac{\partial}{\partial x^{\rho}} (\sqrt{-g} F^{\rho\sigma}{}_{\sigma}) + \sqrt{-g} L_{\text{G-den}},$$

where $L_{\text{G-den}}$ is expressed by (11). According to the above formula and the discussion in §2.2, $\sqrt{-g} R$ changes according to

$$\sqrt{-g} R \to \sqrt{-g} R - 2 \frac{\partial W^{\mu}(x)}{\partial x^{\sigma}} + 2 \frac{\partial W^{\mu}(x)}{\partial x^{\sigma}} = \sqrt{-g} R$$

under T-2, where $W^{\mu}(x)$ is given by (22), we see that $\sqrt{-g} R$ is invariable. And, further, it can be proved easily that $\sqrt{-g} = \det e^{a}_{\mu}$ is also invariable under T-2, $R$ is thus invariable under T-2. This can be seemed as a proof of the conclusion that $R$ is a scalar in the local frame coordinate system.

From (23) it is foreseeable that the Einstein equation $R^{\mu\nu} - \frac{1}{2} g^{\mu\nu} R = \frac{8\pi G}{c^4} T^{\mu\nu}$ can be expressed by $F_{\mu\nu\lambda}$. For the sake of brevity, we define

$$S^{\mu\nu\lambda} = F^{\mu\nu\lambda} + g^{\lambda\mu} F^{\nu\sigma}{}_{\sigma} - g^{\lambda\nu} F^{\mu\sigma}{}_{\sigma},$$

(24)

$S^{\mu\nu\lambda}$ is a three-index tensor for the global manifold coordinate system (when the local frame coordinate system is fixed) and can by proved to be antisymmetric in the first pair of indices:

$$S^{\mu\nu\lambda} = -S^{\nu\mu\lambda}.$$

(25)

From (24) we have

$$S^{\mu\lambda}{}_{\lambda} = -2 F^{\mu\lambda}{}_{\lambda};$$

And, thus, using $-\frac{1}{2} S^{\mu\lambda}{}_{\lambda}$ to substitute $F^{\mu\lambda}{}_{\lambda}$ in (24), we have

$$F^{\mu\nu\lambda} = S^{\mu\nu\lambda} + \frac{1}{2} g^{\lambda\mu} S^{\nu\sigma}{}_\sigma - \frac{1}{2} g^{\lambda\nu} S^{\mu\sigma}{}_\sigma .\qquad(26)$$

According to the formulas from (23) to (26) we can prove the Einstein tensor

$$R^{\mu\nu} - \frac{1}{2} g^{\mu\nu} R = S^{\mu\lambda\nu}{}_{;\lambda} + s^{\mu\nu} = \frac{1}{\sqrt{-g}} \frac{\partial}{\partial x^\lambda} \left( \sqrt{-g} S^{\mu\lambda\nu} \right) + \Gamma^\nu_{\rho\sigma} S^{\mu\rho\sigma} + s^{\mu\nu} ,$$

$$s^{\mu\nu} = F^{\rho\mu\nu} F_{\rho\sigma}{}^\sigma - F^{\rho\mu\sigma} F_{\rho\sigma}{}^\nu - \frac{1}{2} g^{\mu\nu} ( F^{\rho\sigma}{}_\sigma F_{\rho\lambda}{}^\lambda - F^{\rho\sigma\lambda} F_{\rho\lambda\sigma} ) \qquad(27)$$

$$= S^{\rho\mu\nu} S_{\rho\sigma}{}^\sigma + S^{\mu\rho\sigma} S_{\rho\sigma}{}^\nu + \frac{1}{2} S^{\mu\nu\rho} S_{\rho\sigma}{}^\sigma + \frac{1}{2} g^{\mu\nu} S^{\rho\sigma\lambda} S_{\rho\lambda\sigma} - \frac{1}{4} g^{\mu\nu} S^{\rho\sigma}{}_\sigma S_{\rho\lambda}{}^\lambda .$$

The Einstein equation can be written in the following two equivalent forms:

$$S^{\mu\lambda\nu}{}_{;\lambda} + s^{\mu\nu} = \frac{8\pi G}{c^4} T^{\mu\nu} ,$$

$$\frac{\partial}{\partial x^\lambda} \left( \sqrt{-g} S^{\mu\lambda\nu} \right) + \sqrt{-g} \Gamma^\nu_{\rho\sigma} S^{\mu\rho\sigma} + \sqrt{-g} s^{\mu\nu} = \frac{8\pi G}{c^4} \sqrt{-g} T^{\mu\nu} .$$

According to the above second equivalent form of the Einstein equation and using (25) we have

$$\frac{\partial}{\partial x^\mu} \left\{ \sqrt{-g} \left[ T^{\mu\nu} + \frac{c^4}{8\pi G} ( \Gamma^\nu_{\rho\sigma} S^{\rho\mu\sigma} - s^{\mu\nu} ) \right] \right\} = \frac{c^4}{8\pi G} \frac{\partial}{\partial x^\mu} \frac{\partial}{\partial x^\lambda} \left( \sqrt{-g} S^{\mu\lambda\nu} \right) = 0 .$$

Further, according to the above second equivalent form of the Einstein equation we can prove that

$$S^{\mu\nu\hat{\alpha}} = S^{\mu\nu\lambda} e_\lambda^{\hat{\alpha}}$$

satisfies

$$\frac{\partial}{\partial x^\nu} \left( \sqrt{-g} S^{\mu\nu\hat{\alpha}} \right) - \sqrt{-g} \tilde{s}^{\mu\nu} e_\nu^{\hat{\alpha}} = \frac{8\pi G}{c^4} \sqrt{-g} T^{\mu\nu} e_\nu^{\hat{\alpha}} ,$$

$$\tilde{s}^{\mu\nu} = S^{\mu\rho\sigma} F^\nu{}_{\sigma\rho} - s^{\mu\nu}$$

$$= -F^{\mu\rho\sigma} F_{\rho\sigma}{}^\nu - F^{\mu\nu\rho} F_{\rho\sigma}{}^\sigma - F^{\rho\mu\nu} F_{\rho\sigma}{}^\sigma$$

$$+ F^{\mu\rho\sigma} F^\nu{}_{\sigma\rho} - F^{\mu\rho}{}_\rho F^{\nu\sigma}{}_\sigma + \frac{1}{2} g^{\mu\nu} ( F^{\rho\sigma}{}_\sigma F_{\rho\lambda}{}^\lambda - F^{\rho\sigma\lambda} F_{\rho\lambda\sigma} )$$

$$= S^{\mu\rho\sigma} S^\nu{}_{\sigma\rho} - S^{\rho\mu\nu} S_{\rho\sigma}{}^\sigma - S^{\mu\rho\sigma} S_{\rho\sigma}{}^\nu$$

$$- \frac{1}{2} S^{\mu\rho}{}_\rho S^{\nu\sigma}{}_\sigma - \frac{1}{2} g^{\mu\nu} S^{\rho\sigma\lambda} S_{\rho\lambda\sigma} + \frac{1}{4} g^{\mu\nu} S^{\rho\sigma}{}_\sigma S_{\rho\lambda}{}^\lambda .$$

This is the third equivalent form of the Einstein equation. We see that the Christoffel symbol $\Gamma^\nu_{\rho\sigma}$ vanishes in the expression of $\tilde{s}^{\mu\nu}$. Notice

$$S^{\mu\nu\hat{\alpha}} = -S^{\nu\mu\hat{\alpha}} ,$$

according to the above third equivalent form of the Einstein equation we have

$$\frac{\partial}{\partial x^\mu} \left[ \sqrt{-g} \left( T^{\mu\hat{\alpha}} + \frac{c^4}{8\pi G} \tilde{s}^{\mu\hat{\alpha}} \right) \right] = \frac{c^4}{8\pi G} \frac{\partial}{\partial x^\mu} \frac{\partial}{\partial x^\lambda} \left( \sqrt{-g} S^{\mu\lambda\hat{\alpha}} \right) = 0 ,$$

$$T^{\mu\hat{\alpha}} + \frac{c^4}{8\pi G}\widetilde{s}^{\mu\hat{\alpha}} = e^{\hat{\alpha}}_{\nu}\left(T^{\mu\nu} + \frac{c^4}{8\pi G}\widetilde{s}^{\mu\nu}\right).$$

We can rewrite $L_{\text{G-den}}$ given by (11) in terms of $S_{\mu\nu\lambda}$:

$$L_{\text{G-den}} = \frac{1}{2}S^{\mu\nu}{}_{\nu}S_{\mu\lambda}{}^{\lambda} - S^{\mu\nu\lambda}S_{\mu\lambda\nu}.$$

## 3  The Dirac equation in curved spacetime and the action of TGESF

The Dirac equation in curved spacetime[13]

$$\left(i\gamma^{\mu}(x)\frac{\partial}{\partial x^{\mu}} - \frac{1}{4}\gamma^{\hat{\lambda}}r_{\hat{\alpha}\hat{\beta}\hat{\chi}}\sigma^{\hat{\alpha}\hat{\beta}} - \frac{e}{\hbar}\gamma^{\mu}(x)A_{\mu}(x) - \frac{mc}{\hbar}\right)\Psi(x) = 0$$

can be expressed by $F_{\mu\nu\lambda}$ as well:

$$\left(i\gamma^{\mu}(x)\frac{\partial}{\partial x^{\mu}} - \frac{1}{4}\gamma^{\mu}(x)F_{\alpha\beta\mu}\sigma^{\alpha\beta}(x) - \frac{e}{\hbar}\gamma^{\mu}(x)A_{\mu}(x) - \frac{mc}{\hbar}\right)\Psi(x) = 0, \quad (28)$$

where $r_{\hat{\alpha}\hat{\beta}\hat{\gamma}}$ is the Ricci's coefficients of rotation, $A_{\mu}(x)$ is the 4-dimensional electromagnetic potential;

$$\{\gamma^{\hat{\alpha}}, \gamma^{\hat{\beta}}\} = -2\eta^{\hat{\alpha}\hat{\beta}}, \sigma^{\hat{\alpha}\hat{\beta}} = \frac{i}{2}[\gamma^{\hat{\alpha}}, \gamma^{\hat{\beta}}];$$

$$\gamma^{\mu}(x) = \gamma^{\hat{\alpha}}e^{\mu}_{\hat{\alpha}}, \{\gamma^{\mu}(x), \gamma^{\nu}(x)\} = 2g^{\mu\nu}, \sigma^{\mu\nu}(x) = \frac{i}{2}[\gamma^{\mu}(x), \gamma^{\nu}(x)]. \quad (29)$$

The Dirac equation (28) can be derived by the following action

$$L_{\text{D}} = \int L_{\text{D-den}}\sqrt{-g}\,d^4x = \int L_{\text{D-den}}\left(\det e^{\hat{\alpha}}_{\mu}\right)d^4x,$$

$$L_{\text{D-den}} = \overline{\Psi}(x)\left(i\hbar c\gamma^{\mu}(x)\frac{\partial}{\partial x^{\mu}} - \frac{1}{4}\hbar c\gamma^{\mu}(x)F_{\alpha\beta\mu}\sigma^{\alpha\beta}(x) - ec\gamma^{\mu}(x)A_{\mu}(x) - mc^2\right)\Psi(x), \quad (30)$$

where $\overline{\Psi}(x)$ is the *Pauli conjugate* of $\Psi(x)$. The total action of the interaction of gravitational, electromagnetic and spinor field is

$$L_{\text{GES}} = \int L_{\text{GES-den}}\sqrt{-g}\,d^4x = \int L_{\text{GES-den}}\left(\det e^{\hat{\alpha}}_{\mu}\right)d^4x,$$

$$L_{\text{GES-den}} = \frac{c^4}{8\pi G}L_{\text{G-den}} + L_{\text{EM-den}} + L_{\text{D-den}}, \quad (31)$$

where $L_{\text{G-den}}$ and $L_{\text{D-den}}$ are expressed by (11) and (30), respectively; and the Lagrange density of electromagnetic field

$$L_{\text{EM-den}} = -\frac{1}{4\mu_0}F_{\mu\nu}F^{\mu\nu}, F_{\mu\nu} = A_{\nu,\mu} - A_{\mu,\nu}, F^{\mu\nu} = g^{\mu\rho}g^{\nu\sigma}F_{\rho\sigma}. \quad (32)$$

It is obvious that the both Dirac equation and action given by (28) and (30) respectively are invariable under T-1. Contrarily, under T-2, assuming that the wave function $\Psi(x)$ changes

according to

$$\Psi(x) \to S(x)\Psi(x),$$

where $S(x)$ satisfies

$$S(x)\Lambda^{\hat{\alpha}}_{\hat{\beta}}(x)\gamma^{\hat{\beta}} S^{-1}(x) = \gamma^{\hat{\alpha}}. \tag{33}$$

Using (29), (15) and (16), the left-hand of the Dirac equation given by (28) changes according to

$$\left( i\gamma^{\mu}(x)\frac{\partial}{\partial x^{\mu}} - \frac{1}{4}\gamma^{\mu}(x)F_{\alpha\beta\mu}\sigma^{\alpha\beta}(x) - \frac{e}{\hbar}\gamma^{\mu}(x)A_{\mu}(x) - \frac{mc}{\hbar} \right)\Psi(x)$$

$$\to \left( i\gamma^{\mu}(x)\frac{\partial}{\partial x^{\mu}} - \frac{1}{4}\gamma^{\mu}(x)F_{\alpha\beta\mu}\sigma^{\alpha\beta}(x) - \frac{e}{\hbar}\gamma^{\mu}(x)A_{\mu}(x) - \frac{mc}{\hbar} \right)\Psi(x)$$

$$+ \gamma^{\mu}(x)\left[ iS^{-1}(x)\frac{\partial S(x)}{\partial x^{\mu}} - \frac{1}{4}\Lambda^{\hat{\alpha}}_{\hat{\beta}}(x)\frac{\partial \overline{\Lambda}^{\hat{\gamma}}_{\hat{\alpha}}(x)}{\partial x^{\mu}}\eta_{\hat{\gamma}\hat{\chi}}\sigma^{\hat{\beta}\hat{\chi}} \right]\Psi(x).$$

We can prove that $S(x)$ satisfying (33) satisfies

$$iS^{-1}(x)\frac{\partial S(x)}{\partial x^{\mu}} = \frac{1}{4}\Lambda^{\hat{\alpha}}_{\hat{\beta}}(x)\frac{\partial \overline{\Lambda}^{\hat{\gamma}}_{\hat{\alpha}}(x)}{\partial x^{\mu}}\eta_{\hat{\gamma}\hat{\chi}}\sigma^{\hat{\beta}\hat{\chi}}. \tag{34}$$

For this purpose, it is enough to consider the case of the local infinitesimal Lorentz transformations

$$\Lambda^{\hat{\alpha}}_{\hat{\beta}}(x) = \delta^{\hat{\alpha}}_{\hat{\beta}} + \Delta\varpi^{\hat{\alpha}}_{\hat{\beta}}(x), \quad \overline{\Lambda}^{\hat{\alpha}}_{\hat{\beta}}(x) = \delta^{\hat{\alpha}}_{\hat{\beta}} - \Delta\varpi^{\hat{\alpha}}_{\hat{\beta}}(x), \quad \Delta\varpi_{\hat{\alpha}\hat{\beta}}(x) = \eta_{\hat{\beta}\hat{\gamma}}\Delta\varpi^{\hat{\gamma}}_{\hat{\alpha}}(x).$$

In this case[14],

$$S(x) = 1 - \frac{i}{4}\sigma^{\hat{\alpha}\hat{\beta}}\Delta\varpi_{\hat{\alpha}\hat{\beta}}(x);$$

Substituting the above expressions to the right- and left-hand of (34), respectively, we therefore see that (34) holds, that means that the Dirac equation given by (28) is invariable under T-2.

Simply setting $\overline{\Psi}(x) = \Psi^{+}(x)\beta$, where $\beta = \gamma^0 = \begin{pmatrix} I & 0 \\ 0 & -I \end{pmatrix}$ in the standard representation of the Dirac matrices, and notice that $S(x)$ satisfying (33) satisfies

$$S^{+}(x)\beta S(x) = \beta,$$

we can prove that $L_{\text{D-den}}$ given by (30) is also invariable under T-2.

One can try to add other terms into (31), however, (31) itself has already been variable neither under T-1 nor the T-2, that means that the theory given by (31) can be as a independent theory. It is obvious that the theory given by (31) is *minimal* theory on the interaction of gravitational, electromagnetic and spinor field.

The interactional term of gravitational field and spinor particle in (30) $-\frac{1}{4}\hbar c F_{\alpha\beta\mu}\overline{\Psi}(x)\gamma^{\mu}(x)\sigma^{\alpha\beta}(x)\Psi(x)$ is independent of some parameters of spinor particle, e.g., the mass $m$ of the particle; in other words, the interaction of gravitational field and spinor particle given by (30) also satisfies *the equivalence principle*.

Now we summarize the whole transformations with the gauge transformations (T-3) and the corresponding characters of transformations of $L_{\text{GES}}$ in Table 1.

**Tab. 1  The characters of transformations of $L_{\text{GES}}$**

|  | T-1: $x^\mu = x^\mu(\tilde{x}^\nu)$ | T-2: $\mathrm{d}\xi^{\hat{\alpha}} = \Lambda^{\hat{\alpha}}_{\hat{\beta}}(x)\mathrm{d}\tilde{\xi}^{\hat{\beta}}$ | T-3: gauge transformations |
|---|---|---|---|
| $e^{\hat{\alpha}}_\mu$ | $e^{\hat{\alpha}}_\mu = \dfrac{\partial \tilde{x}^\nu}{\partial x^\mu}\tilde{e}^{\hat{\alpha}}_\nu$ | $e^{\hat{\alpha}}_\mu = \Lambda^{\hat{\alpha}}_{\hat{\beta}}(x)\tilde{e}^{\hat{\beta}}_\mu$ | Invariable |
| $A_\mu$ | $A_\mu = \dfrac{\partial \tilde{x}^\nu}{\partial x^\mu}\tilde{A}_\nu$ | Invariable | $A_\mu(x) = \tilde{A}_\mu(x) - \dfrac{\hbar}{e}\dfrac{\partial \theta(x)}{\partial x^\mu}$ |
| $\Psi(x)$ | $\Psi(x) = \Psi(x(\tilde{x}))$ | $\Psi(x) = S(x)\tilde{\Psi}(x)$ | $\Psi(x) = \mathrm{e}^{\mathrm{i}\theta(x)}\tilde{\Psi}(x)$ |
| $L_{\text{GES}}$ | Invariable | $L_{\text{GES}} = \tilde{L}_{\text{GES}} + 2\dfrac{c^4}{8\pi G}\int \dfrac{\partial W^\mu(x)}{\partial x^\mu}\mathrm{d}^4 x$ | Invariable |

## 4  Gravitational field being as a gauge field

### 4.1  The introduction of gravitational field in terms of the principle of gauge invariance based on the Dirac equation

Gravitational field can be introduced in terms of the method of introducing Yang-Mills gauge field based on the Dirac equation. Let us start from the Dirac equation in flat spacetime in which $g_{\mu\nu} = \eta_{\mu\nu}$:

$$\left(\mathrm{i}\gamma^{\hat{\alpha}}\frac{\partial}{\partial X^{\hat{\alpha}}} - \frac{mc}{\hbar}\right)\psi(X) = 0,$$

where $\gamma^{\hat{\alpha}}$ are constant matrices. For the purpose obtaining a corresponding equation that holds in an arbitrary coordinate system, after an arbitrary coordinate transformation

$$X^{\hat{\alpha}} = X^{\hat{\alpha}}(x^\lambda), \tag{35}$$

we obtain the Dirac equation in an arbitrary noninertial system

$$\left(\mathrm{i}\gamma^{\hat{\alpha}}e^\mu_{\hat{\alpha}}\frac{\partial}{\partial x^\mu} - \frac{mc}{\hbar}\right)\Psi(x) = 0,$$

$$e^\mu_{\hat{\alpha}} = e^\mu_{\hat{\alpha}}(x) = \frac{\partial x^\mu}{\partial X^{\hat{\alpha}}}, \Psi(x) = \psi(X(x)). \tag{36}$$

Notice that the effect of the original flat spacetime coordinate system $\{X^{\hat{\alpha}}\}$ does not vanish after the coordinate transformation (35), it still works by $e_{\hat{\alpha}}^{\mu} = \dfrac{\partial x^{\mu}}{\partial X^{\hat{\alpha}}}$.

The similar case occurs on other fields as well. For instance, after the coordinate transformation (35), the Klein-Gordon equation

$$\left(\eta^{\hat{\alpha}\hat{\beta}} \frac{\partial}{\partial X^{\hat{\alpha}}} \frac{\partial}{\partial X^{\hat{\beta}}} - \frac{m^2 c^2}{\hbar^2}\right) \varphi(X) = 0,$$

changes to

$$\left(g^{\mu\nu} \frac{\partial}{\partial x^{\mu}} \frac{\partial}{\partial x^{\nu}} - \frac{m^2 c^2}{\hbar^2}\right) \Phi(x) = 0,$$

$$g^{\mu\nu} = \eta^{\hat{\alpha}\hat{\beta}} \frac{\partial x^{\mu}}{\partial X^{\hat{\alpha}}} \frac{\partial x^{\nu}}{\partial X^{\hat{\beta}}}, \Phi(x) = \varphi(X(x)).$$

We see that in the above Klein-Gordon equation in an arbitrary noninertial system, the original flat spacetime coordinate system $\{X^{\hat{\alpha}}\}$ still works by $g^{\mu\nu} = \eta^{\hat{\alpha}\hat{\beta}} \dfrac{\partial x^{\mu}}{\partial X^{\hat{\alpha}}} \dfrac{\partial x^{\nu}}{\partial X^{\hat{\beta}}}$.

Now we go back to (36). For the purpose removing fully the effect of the special coordinate system $\{X^{\hat{\alpha}}\}$, we do another *arbitrary* coordinate transformation for $\{X^{\hat{\alpha}}\}$

$$X^{\hat{\alpha}} = X^{\hat{\alpha}}(\xi^{\hat{\beta}})|_{(x^{\lambda})}, \tag{37}$$

in (37), the subscript $(x^{\lambda})$ means that the form of the coordinate transformation $X^{\hat{\alpha}} = X^{\hat{\alpha}}(\xi^{\hat{\beta}})$ maybe different at different spacetime point. It is obvious that the effect of the original special coordinate system $\{X^{\hat{\alpha}}\}$ is removed fully after the coordinate transformation (37).

Under the coordinate transformation (37), the quantities in (36) change according to

$$\gamma^{\hat{\alpha}} = \Lambda_{\hat{\beta}}^{\hat{\alpha}}(x) \widetilde{\gamma}^{\hat{\beta}}(x), e_{\hat{\alpha}}^{\mu} = \overline{\Lambda}_{\hat{\alpha}}^{\hat{\beta}}(x) \widetilde{e}_{\hat{\beta}}^{\mu};$$

$$\Lambda_{\hat{\beta}}^{\hat{\alpha}}(x) = \frac{\partial X^{\hat{\alpha}}}{\partial \xi^{\hat{\beta}}}, \overline{\Lambda}_{\hat{\alpha}}^{\hat{\beta}}(x) = \frac{\partial \xi^{\hat{\beta}}}{\partial X^{\hat{\alpha}}}; \Lambda_{\hat{\beta}}^{\hat{\alpha}}(x) \overline{\Lambda}_{\hat{\alpha}}^{\hat{\gamma}}(x) = \delta_{\hat{\beta}}^{\hat{\gamma}}, \Lambda_{\hat{\alpha}}^{\hat{\gamma}}(x) \overline{\Lambda}_{\hat{\beta}}^{\hat{\alpha}}(x) = \delta_{\hat{\beta}}^{\hat{\gamma}}.$$

$\Lambda_{\hat{\beta}}^{\hat{\alpha}} = \Lambda_{\hat{\beta}}^{\hat{\alpha}}(x)$ is a function of the coordinate $\{x^{\mu}\}$, that means that $\Lambda_{\hat{\beta}}^{\hat{\alpha}}$ at different spacetime point maybe different. And, further, (36) changes to the form

$$\left(i\widetilde{\gamma}^{\hat{\alpha}}(x) \widetilde{e}_{\hat{\alpha}}^{\mu} \frac{\partial}{\partial x^{\mu}} - \frac{mc}{\hbar}\right) \Psi(x) = 0. \tag{38}$$

Comparing the above equation with (36), we see that the form of (36) has been changed. However, if we try to change $\widetilde{\gamma}^{\hat{\alpha}}(x)$ to the original $\gamma^{\hat{\alpha}}$, then the equation remains in the same form.. For this purpose, we set that $\Psi(x)$ changes according to

$$\Psi(x) = S(x)\widetilde{\Psi}(x),$$

and left-multiplying $S^{-1}(x)$ on (38), we obtain

$$\left( iS^{-1}(x)\widetilde{\gamma}^{\hat{\alpha}}(x)S(x)\widetilde{e}_{\hat{\alpha}}^{\mu}\frac{\partial}{\partial x^{\mu}} + iS^{-1}(x)\widetilde{\gamma}^{\hat{\alpha}}(x)S(x)\widetilde{e}_{\hat{\alpha}}^{\mu}S^{-1}(x)\frac{\partial S(x)}{\partial x^{\mu}} - \frac{mc}{\hbar} \right)\widetilde{\Psi}(x) = 0,$$

and assuming

$$S^{-1}(x)\widetilde{\gamma}^{\hat{\alpha}}(x)S(x) = \gamma^{\hat{\alpha}},$$

this is just (33). Now (38) is in the form

$$\left( i\gamma^{\hat{\alpha}}\widetilde{e}_{\hat{\alpha}}^{\mu}\frac{\partial}{\partial x^{\mu}} + i\gamma^{\hat{\alpha}}\widetilde{e}_{\hat{\alpha}}^{\mu}S^{-1}(x)\frac{\partial S(x)}{\partial x^{\mu}} - \frac{mc}{\hbar} \right)\widetilde{\Psi}(x) = 0.$$

We see, although $\widetilde{\gamma}^{\hat{\alpha}}(x)$ are changed to the original $\gamma^{\hat{\alpha}}$, an extra term $i\gamma^{\hat{\alpha}}\widetilde{e}_{\hat{\alpha}}^{\mu}S^{-1}(x)\frac{\partial S(x)}{\partial x^{\mu}}\widetilde{\Psi}(x)$ appears. For removing this extra term, we have to add a term $F$ into the original equation (36), which thus becomes

$$\left( i\gamma^{\hat{\alpha}}e_{\hat{\alpha}}^{\mu}\frac{\partial}{\partial x^{\mu}} + F - \frac{mc}{\hbar} \right)\Psi(x) = 0.$$

We ask $F$ is invariable under the general coordinate transformation $x^{\mu} = x^{\mu}(\widetilde{x}^{\nu})$, and changes according to

$$S^{-1}(x)FS(x) = \widetilde{F} + i\gamma^{\hat{\alpha}}\widetilde{e}_{\hat{\alpha}}^{\mu}S^{-1}(x)\frac{\partial S(x)}{\partial x^{\mu}} \tag{39}$$

under the coordinate transformation (37). According to the discussion in the Section 3, we know that

$$F = -\frac{1}{4}\gamma^{\mu}(x)F_{\alpha\beta\mu}\sigma^{\alpha\beta}(x) \tag{40}$$

satisfies (39).

Now that $F_{\mu\nu\lambda}$ appears in the theory, we have to add the action $L(F_{\mu\nu\lambda})$ about $F_{\mu\nu\lambda}$ itself. We ask the action $L(F_{\mu\nu\lambda})$ is variable neither under the coordinate transformation $x^{\mu} = x^{\mu}(\widetilde{x}^{\nu})$ nor (37). According to the anterior discussion we know that the action given by (6) and (11) about $F_{\mu\nu\lambda}$ itself is invariable under the coordinate transformation $x^{\mu} = x^{\mu}(\widetilde{x}^{\nu})$, and the change under (37) is only an integral of a total derivative that is dynamically irrelevant, this is acceptable.

**4.2　The gravitational potential**

So far, the metric tensor $g_{\mu\nu}$ has being regarded as the basic variable in large numbers of works. It is usually split two parts: $g_{\mu\nu} = \overline{g}_{\mu\nu} + h_{\mu\nu}$, where the $c$-number quantity $\overline{g}_{\mu\nu}$ is as the background field, one can choose simply[7] $\overline{g}_{\mu\nu} = \eta_{\mu\nu}$, this is allowable in terms of the equivalence principle, and $h_{\mu\nu}$ is regarded as *the gravitational potential*.

However, we have to choose the tetrad field $e_\mu^{\hat{\alpha}}$ as the basic variable here, because only $e_\mu^{\hat{\alpha}}$ can expresses completely $F_{\mu\nu\lambda}$ by which the Lagrange density of the gravitational field $L_{\text{G-den}}$ expressed by (11) that we use here is constructed.

If $e_\mu^{\hat{\alpha}}$ satisfies $e_{\mu,\nu}^{\hat{\alpha}} = e_{\nu,\mu}^{\hat{\alpha}}$ then from (10), $F_{\mu\nu\lambda} = 0$, and, further, from (23), $R_{\rho\nu\sigma}^\mu = 0$, this is so called *the condition of integrability*. As well known[12], in this case we can find out a coordinate transformation $x^\mu = x^\mu(X^\lambda)$ so as to $g_{\mu\nu}(X^\lambda) = \eta_{\mu\nu}$ in the special coordinate system $\{X^\lambda\}$.

According to the above condition of integrability, real gravity exists if and only if $e_{\mu,\nu}^{\hat{\alpha}} \neq e_{\nu,\mu}^{\hat{\alpha}}$. In this case we can split $e_\mu^{\hat{\alpha}}(x^\lambda)$ into the trivial part $\bar{h}_\mu^{\hat{\alpha}}(x^\lambda)$ and the nontrivial part $h_\mu^{\hat{\alpha}}(x^\lambda)$: $e_\mu^{\hat{\alpha}}(x^\lambda) = \bar{h}_\mu^{\hat{\alpha}}(x^\lambda) + h_\mu^{\hat{\alpha}}(x^\lambda)$ in the coordinate system $\{x^\lambda\}$, where

$$\bar{h}_\mu^{\hat{\alpha}}(x^\lambda) = \frac{\partial X^{\hat{\alpha}}}{\partial x^\mu} \tag{41}$$

satisfies

$$\bar{h}_{\mu,\nu}^{\hat{\alpha}}(x^\lambda) = \frac{\partial^2 X^{\hat{\alpha}}}{\partial x^\nu \partial x^\mu} = \frac{\partial^2 X^{\hat{\alpha}}}{\partial x^\mu \partial x^\nu} = \bar{h}_{\nu,\mu}^{\hat{\alpha}}(x^\lambda),$$

where $X^{\hat{\alpha}} = X^{\hat{\alpha}}(x^\lambda)$ is an arbitrary function. Simply, we can choose

$$\bar{h}_\mu^{\hat{\alpha}}(x^\lambda) = \delta_\mu^{\hat{\alpha}}, \tag{42}$$

this is allowable in terms of the equivalence principle.

According to the above analyze we see that the nontrivial part

$$h_\mu^{\hat{\alpha}}(x^\lambda) = e_\mu^{\hat{\alpha}}(x^\lambda) - \bar{h}_\mu^{\hat{\alpha}}(x^\lambda) \tag{43}$$

of $e_\mu^{\hat{\alpha}}(x^\lambda)$ denote real gravitational field, we therefore can regard $h_\mu^{\hat{\alpha}}(x^\lambda)$ as the gravitational potential.

Differing from the Yang-Mill field for which it is the potential $A_\mu^a$ that interacts with the Dirac field, from (10), (43) and (28) we see that, for gravitational field, it is the gravitational *field-strength* $F_{\mu\nu\lambda}$ including the (first) derivative of the gravitational potential $h_\mu^{\hat{\alpha}}$, but not only the gravitational *potential* $h_\mu^{\hat{\alpha}}$, that interacts with the Dirac field, this result arises from the equivalence principle. Because, according to the equivalence principle, gravity at a local area can be vanished, the derivative of $h_\mu^{\hat{\alpha}}$ means the effect of the gravitational field that cannot be vanished at a local area[12].

### 4.3 The groups of coordinate transformations of gravitational field

We have shown that how to introduce gravitational field based on the principle of gauge invariance under the two coordinate transformations: T-1 and T-2. As well known, the both T-1 and T-2 construct the groups $U_{\text{T-1}}$ and $U_{\text{T-2}}$, respectively. An arbitrary element $u$ in the groups $U_{\text{T-1}}$ is

$x^\mu = x^\mu(\tilde{x}^\lambda)$ of which the manner acting on $e_\mu^{\hat{\alpha}}$ is

$$^u(e_\mu^{\hat{\alpha}}) = \tilde{e}_\mu^{\hat{\alpha}} = \frac{\partial x^\nu}{\partial \tilde{x}^\mu} e_\nu^{\hat{\alpha}}.$$

The manner that $u$ acts on other quantities in the theory can be obtained by analogous method. Accordingly, an arbitrary element $\hat{u}$ in the group $U_{\text{T-2}}$ is $\Lambda_{\hat{\beta}}^{\hat{\alpha}}(x)$, according to (15), the manner that $\hat{u}$ acts on $e_\mu^{\hat{\alpha}}$ is

$$^{\hat{u}}(e_\mu^{\hat{\alpha}}) = \tilde{e}_\mu^{\hat{\alpha}} = \overline{\Lambda}_{\hat{\beta}}^{\hat{\alpha}}(x) e_\mu^{\hat{\beta}}.$$

The manner that $\hat{u}$ acts on other quantities in the theory can be obtained by analogous method.

However, we have to consider the case that the both coordinate transformations T-1 and T-2 concur, that means that we must compose the two groups $U_{\text{T-1}}$ and $U_{\text{T-2}}$ to a bigger group $U_{(\text{T-1})\wedge(\text{T-2})}$, and the manner that an arbitrary element $u\hat{u}$ in the group $U_{(\text{T-1})\wedge(\text{T-2})}$ acts on $e_\mu^{\hat{\alpha}}$ is

$$^u(^{\hat{u}}(e_\mu^{\hat{\alpha}}(x))) = ^u(\overline{\Lambda}_{\hat{\beta}}^{\hat{\alpha}}(x) e_\mu^{\hat{\beta}}(x)) = \frac{\partial x^\nu}{\partial \tilde{x}^\mu} \overline{\Lambda}_{\hat{\beta}}^{\hat{\alpha}}(\tilde{x}) e_\nu^{\hat{\beta}}(\tilde{x})$$

$$= \overline{\Lambda}_{\hat{\beta}}^{\hat{\alpha}}(\tilde{x}) \left( \frac{\partial x^\nu}{\partial \tilde{x}^\mu} e_\nu^{\hat{\beta}}(\tilde{x}) \right) = ^{\hat{u}}\left( \frac{\partial x^\nu}{\partial \tilde{x}^\mu} e_\nu^{\hat{\beta}}(\tilde{x}) \right) = ^{\hat{u}}(^u(e_\mu^{\hat{\alpha}}(x))),$$

i.e., $u\hat{u} = \hat{u}u$. The manner that $u\hat{u}$ acts on other quantities in the theory can be obtained by analogous method.

Especially, for the infinitesimal group of coordinate transformations $\Delta U_{(\text{T-1})\wedge(\text{T-2})}$ in which $\Delta u$ is $x^\mu = x^\mu + \Delta \zeta^\mu(\tilde{x}^\lambda)$ and $\Delta \hat{u}$ is $\Lambda_{\hat{\beta}}^{\hat{\alpha}}(x) = \delta_{\hat{\beta}}^{\hat{\alpha}} + \Delta \varpi_{\hat{\beta}}^{\hat{\alpha}}(x)$, we have

$$^{\Delta u \Delta \hat{u}}(e_\mu^{\hat{\alpha}}(x)) = \tilde{e}_\mu^{\hat{\alpha}}(\tilde{x}) = \overline{\Lambda}_{\hat{\beta}}^{\hat{\alpha}}(\tilde{x}) \left( \frac{\partial x^\nu}{\partial \tilde{x}^\mu} e_\nu^{\hat{\beta}}(\tilde{x}) \right)$$

$$= e_\mu^{\hat{\alpha}}(\tilde{x}) - \eta^{\hat{\alpha}\hat{\gamma}} \Delta \varpi_{\hat{\beta}\hat{\gamma}}(\tilde{x}) e_\mu^{\hat{\beta}}(\tilde{x}) + \frac{\partial \Delta \zeta^\nu(\tilde{x}^\lambda)}{\partial \tilde{x}^\mu} e_\nu^{\hat{\alpha}}(\tilde{x}),$$

(44)

where $\Delta \varpi_{\hat{\beta}\hat{\gamma}} = \eta_{\hat{\alpha}\hat{\gamma}} \Delta \varpi_{\hat{\beta}}^{\hat{\alpha}}$, $\Delta \varpi_{\hat{\beta}\hat{\gamma}} = -\Delta \varpi_{\hat{\gamma}\hat{\beta}}$; and the second order infinitesimal quantities in (44) are dropped. The manner that $\Delta u \Delta \hat{u}$ acts on other quantities in the theory can be obtained by analogous method.

There are ten independent parameters in $\Delta U_{(\text{T-1})\wedge(\text{T-2})}$. We choose

$$\varepsilon^{(k)} = \Delta \varpi_{\hat{0}\hat{k}} \ (k = 1, 2, 3),$$
$$\varepsilon^{(4)} = \Delta \varpi_{\hat{1}\hat{2}}, \ \varepsilon^{(5)} = \Delta \varpi_{\hat{1}\hat{3}}, \ \varepsilon^{(6)} = \Delta \varpi_{\hat{2}\hat{3}},$$
$$\varepsilon^{(\mu+7)} = \Delta \zeta^\mu \ (\mu = 0, 1, 2, 3)$$

(45)

to be as the ten independent parameters.

## 4.4 The constraint conditions for $h_\mu^{\hat{\alpha}}(x^\lambda)$

According to the discussion of §4.2, $h_\mu^{\hat{\alpha}}(x^\lambda)$ which has 16 componentsis now used as the gravitational potential of the theory; from the discussion of §4.3 we see that the theory is invariable under $\Delta U_{(T-1)\wedge(T-2)}$ which has ten independent parameters, we therefore have to add ten constraint conditions.

In fact, in traditional general relativity we know that if we want to express $e_\mu^{\hat{\alpha}}$ which has 16 components in terms of the metric tensor $g_{\mu\nu}$ which has 10 components, we have to add six constraint conditions; and, further, there are only six independent equations in the Einstein equation[12], we have to add four extra equations as the constraint conditions for determining ten components of the metric tensor $g_{\mu\nu}$. So we have to add ten constraint conditions on $e_\mu^{\hat{\alpha}}$ in all.

For the ten constraint conditions, *existence*, *completeness* and *consistency* of a group of constraint conditions are necessary. The existence means that there are special element $u_0 \hat{u}_0$ in the group $U_{(T-1)\wedge(T-2)}$, i.e., there are two special coordinate transformations $x^\mu = x^\mu(\tilde{x}^\lambda)$ and $d\xi^{\hat{\alpha}} = \Lambda_{\hat{\beta}}^{\hat{\alpha}}(x) d\tilde{\xi}^{\hat{\beta}}$, to make that the given ten constraint conditions hold; the completeness means that the given ten constraint conditions can take out completely the nonphysical degrees of freedom, the consistency means that there is not any conflict between the given ten constraint conditions.

In spite of some known constraint conditions, e.g., the Prentki gauge[7] $\sum_{i=1}^{3}\frac{\partial h_{i\nu}}{\partial x^i}=0$ ( $\nu = 0,1,2,3$ ) for $h_{\mu\nu} = g_{\mu\nu} - \eta_{\mu\nu}$ which much like the Coulomb gauge in quantum-electrodynamics, we choose the following ten equations as the constraint conditions:

$$e_\alpha^{\hat{\alpha}} = 1,\ \alpha = 0,1,2,3;\ e_\mu^{\hat{\alpha}} = 0,\ \alpha > \mu,\ \alpha, \mu = 0,1,2,3$$

namely,

$$f^{(\alpha+1)} = e_\alpha^{\hat{\alpha}} - 1 = 0\ (\alpha = 0,1,2,3),\ f^{(k+4)} = e_0^{\hat{k}} = 0\ (k = 1,2,3),$$
$$f^{(8)} = e_1^{\hat{2}} = 0,\ f^{(9)} = e_1^{\hat{3}} = 0,\ f^{(10)} = e_2^{\hat{3}} = 0.$$

(46)

Notice that although (46) shows that the ten constraint conditions act on $e_\mu^{\hat{\alpha}}$, they act on actually the gravitational potential $h_\mu^{\hat{\alpha}}$ by (43), (42) or (41).

All the discussions on the physical reason choosing (46), the existence, completeness and consistency of (46) are given in the Appendix. Here we only point out a character of (46) and two concludes arising from the character.

From the discussion in Appendix we know that if (A4) and at the same time (A7) hold, then $F_{\mu\nu\lambda}$ is only a function of $g_{\mu\nu}$ and the first derivative of $g_{\mu\nu}$:

$$F_{\mu\nu\lambda} = F_{\mu\nu\lambda}(g_{\rho\sigma}, \partial_\alpha g_{\rho\sigma}).$$

The following two concludes thus arise from the above character.

① From (11) we know that the Lagrange density of the gravitational field $L_{\text{G-den}} = L_{\text{G-den}}(g_{\rho\sigma}, \partial_\alpha g_{\rho\sigma})$ is likewise a function of $g_{\mu\nu}$ and the first derivative of $g_{\mu\nu}$ now. This is quite different from some earlier covariant quantum theory of gravity in which $\int R\sqrt{-g}\,\mathrm{d}^4 x$ is chosen to be as the action, after dropping some total derivative terms, some second derivative terms of $g_{\mu\nu}$ still remain in the action.

② Because now the both $F_{\mu\nu\lambda}$ and $L_{\text{G-den}}$ are functions of $g_{\mu\nu}$ and the first derivative of $g_{\mu\nu}$, if $F_{\mu\nu\lambda}$ and $L_{\text{G-den}}$ are linearized, then we can obtain a theory of spin-2 field which differs from, although is quite similar to, the Fierz-representation of spin-2 field[4]. Because the Fierz tensor $f_{\mu\nu\lambda}$ has the following characters:

$$f_{\mu\nu\lambda} = -f_{\nu\mu\lambda}, \quad f_{\mu\nu\lambda} + f_{\nu\lambda\mu} + f_{\lambda\mu\nu} = 0.$$

We see, the first of the above two characters is completely the same as (13) and (25); however, neither $F_{\mu\nu\lambda}$ nor $S_{\mu\nu\lambda}$ defined by (10) and (24) respectively has the second character. In fact, from (10) and (24), using (14) we have

$$S_{\mu\nu\lambda} + S_{\nu\lambda\mu} + S_{\lambda\mu\nu} = F_{\mu\nu\lambda} + F_{\nu\lambda\mu} + F_{\lambda\mu\nu} = F_{\mu\nu\lambda} + \eta_{\hat{\alpha}\hat{\beta}} e^{\hat{\alpha}}_\lambda e^{\hat{\beta}}_{\mu,\nu} - \eta_{\hat{\alpha}\hat{\beta}} e^{\hat{\alpha}}_\lambda e^{\hat{\beta}}_{\nu,\mu}$$

$$= \frac{1}{2}\eta_{\hat{\alpha}\hat{\beta}} e^{\hat{\alpha}}_\mu e^{\hat{\beta}}_{\nu,\lambda} - \frac{1}{2}\eta_{\hat{\alpha}\hat{\beta}} e^{\hat{\alpha}}_\nu e^{\hat{\beta}}_{\mu,\lambda} + \frac{1}{2}\eta_{\hat{\alpha}\hat{\beta}} e^{\hat{\alpha}}_\nu e^{\hat{\beta}}_{\lambda,\mu} - \frac{1}{2}\eta_{\hat{\alpha}\hat{\beta}} e^{\hat{\alpha}}_\mu e^{\hat{\beta}}_{\lambda,\nu} + \frac{1}{2}\eta_{\hat{\alpha}\hat{\beta}} e^{\hat{\alpha}}_\lambda e^{\hat{\beta}}_{\mu,\nu} - \frac{1}{2}\eta_{\hat{\alpha}\hat{\beta}} e^{\hat{\alpha}}_\lambda e^{\hat{\beta}}_{\nu,\mu}.$$

So, it seems as if the spin-2 field based on the Fierz tensor is improper for general relativity.

However, we emphasize that the gravitational potential is $h^{\hat{\alpha}}_\mu(x^\lambda) = e^{\hat{\alpha}}_\mu(x^\lambda) - \bar{h}^{\hat{\alpha}}_\mu(x^\lambda)$ but not $h_{\mu\nu} = g_{\mu\nu} - \bar{g}_{\mu\nu}$ in the present paper.

### 4.5 Summarization

We see, the form of general relativity has already been changed to accord with the standard form of the theory of field, especially, to satisfy the principle of gauge invariance. In this theory of gravitational field, the potential of the field is $h^{\hat{\alpha}}_\mu(x^\lambda)$; the field-strength is $F_{\mu\nu\lambda}$ which is a function of $h^{\hat{\alpha}}_\mu(x^\lambda)$ and the first derivative of $h^{\hat{\alpha}}_\mu(x^\lambda)$; the action of the theory are given by (6) and (11); the group of the gauge transformation is that of coordinate transformation $U_{(\text{T-1})\wedge(\text{T-2})}$; finally, the equation of field is one of the three equivalent form of the Einstein equation presented in §2.3.

If we add *the theory on measure*, e.g., for infinitesimal coordinate distance $\mathrm{d}x^\mu$ at spacetime point $x^\mu_P$, the actual measured results of observer is $\mathrm{d}\xi^{\hat{\alpha}} = e^{\hat{\alpha}}_\mu(x^\lambda_P)\mathrm{d}x^\mu$, etc, then we can try to use the theory of gravitational field based on the conception of *field* presented here to replace general relativity based on the geometrical conceptions.

However, could we remove fully the geometrical conceptions in the theory of gravitational field? For instance, could the Riemann curvature tensor be seemed as to be defined by (23) in terms of $F_{\mu\nu\lambda}$ and, is the purpose that one defines the quantity $R^{\mu}_{\rho\nu\sigma}$ merely for convenience? This question will be studied in further.

## 5  Quantization of the pure gravitational field

When we try to quantize the pure gravitational field, if we use the canonical quantization procedure, in spite of the complexity arised from the constraint conditions, what we face at first is the choice of the temporal axis, that means that we have to use a special method, e.g., that of ADM 3+1 decomposition[16], to prescribe the temporal axis in the 4-dimensional spacetime.

However, the Feynman's method of path integral maybe avoids the problem of prescribing the temporal axis in the 4-dimensional spacetime. Because, ① What used in the Feynman's method of path integral is only the Lagrange density, but not the Hamiltonian density. And, neither $L_{\text{G-den}}\left(\det e^{\hat{\alpha}}_{\mu}\right)$ nor $L_{\text{GES-den}}\left(\det e^{\hat{\alpha}}_{\mu}\right)$ asks to prescribe the temporal axis in the 4-dimensional spacetime, where $L_{\text{G-den}}$ and $L_{\text{GES-den}}$ are given by (11) and (31), respectively. ② From the discussion in Appendix we know that the ten constraint conditions given by (46) do not depend upon prescribing the temporal axis in the 4-dimensional spacetime.

When we use the Feynman's method of path integral to quantize the pure gravitational field, we can use the Faddeev-Popov method[15] to deal with an arbitrary group of constraint conditions if only the given group of constraint conditions has the characters of existence, completeness and consistency. Because the set of coordinate transformations $U_{\text{(T-1)}\wedge\text{(T-2)}}$ that leads to the appearance of ten constraint conditions has the character of group, it is just this character that makes the Faddeev-Popov method holds in this case.

For obtaining the Faddeev-Popov determinant[15], after the infinitesimal coordinate transformations in terms of (44), every $f^{(a)}$ $(a = 1, 2, \cdots, 10)$ given by (46) changes to

$$\tilde{f}^{(1)} = \tilde{e}^{\hat{0}}_{0} - 1 = \partial_0 \Delta\zeta^0 + e^{\hat{0}}_1 \partial_0 \Delta\zeta^1 + e^{\hat{0}}_2 \partial_0 \Delta\zeta^2 + e^{\hat{0}}_3 \partial_0 \Delta\zeta^3 ,$$

$$\tilde{f}^{(2)} = \tilde{e}^{\hat{1}}_{1} - 1 = -e^{\hat{0}}_1 \Delta\varpi_{\hat{0}\hat{1}} + e^{\hat{1}}_1 \partial_1 \Delta\zeta^1 + e^{\hat{1}}_2 \partial_1 \Delta\zeta^2 + e^{\hat{1}}_3 \partial_1 \Delta\zeta^3 ,$$

$$\tilde{f}^{(3)} = \tilde{e}^{\hat{2}}_{2} - 1 = -e^{\hat{0}}_2 \Delta\varpi_{\hat{0}\hat{2}} - e^{\hat{1}}_2 \Delta\varpi_{\hat{1}\hat{2}} + e^{\hat{2}}_2 \partial_2 \Delta\zeta^2 + e^{\hat{2}}_3 \partial_2 \Delta\zeta^3 ,$$

$$\tilde{f}^{(4)} = \tilde{e}^{\hat{3}}_{3} - 1 = -e^{\hat{0}}_3 \Delta\varpi_{\hat{0}\hat{3}} - e^{\hat{1}}_3 \Delta\varpi_{\hat{1}\hat{3}} - e^{\hat{2}}_3 \Delta\varpi_{\hat{2}\hat{3}} + e^{\hat{3}}_3 \partial_3 \Delta\zeta^3 ,$$

$$\tilde{f}^{(5)} = \tilde{e}^{\hat{1}}_{0} = -\Delta\varpi_{\hat{0}\hat{1}} + e^{\hat{1}}_1 \partial_0 \Delta\zeta^1 + e^{\hat{1}}_2 \partial_0 \Delta\zeta^2 + e^{\hat{1}}_3 \partial_0 \Delta\zeta^3 ,$$

$$\tilde{f}^{(6)} = \tilde{e}^{\hat{2}}_{0} = -\Delta\varpi_{\hat{0}\hat{2}} + e^{\hat{2}}_2 \partial_0 \Delta\zeta^2 + e^{\hat{2}}_3 \partial_0 \Delta\zeta^3 ,$$

$$\tilde{f}^{(7)} = \tilde{e}^{\hat{3}}_{0} = -\Delta\varpi_{\hat{0}\hat{3}} + e^{\hat{3}}_3 \partial_0 \Delta\zeta^3 ,$$

$$\tilde{f}^{(8)} = \tilde{e}^{\hat{2}}_{1} = -e^{\hat{0}}_1 \Delta\varpi_{\hat{0}\hat{2}} - \Delta\varpi_{\hat{1}\hat{2}} + \partial_1 \Delta\zeta^2 + e^{\hat{2}}_3 \partial_1 \Delta\zeta^3 ,$$

$$\tilde{f}^{(9)} = \tilde{e}^{\hat{3}}_{1} = -e^{\hat{0}}_1 \Delta\varpi_{\hat{0}\hat{3}} - \Delta\varpi_{\hat{1}\hat{3}} + \partial_1 \Delta\zeta^3 ,$$

$$\tilde{f}^{(10)} = \tilde{e}^{\hat{3}}_{2} = -e^{\hat{0}}_2 \Delta\varpi_{\hat{0}\hat{3}} - e^{\hat{1}}_2 \Delta\varpi_{\hat{1}\hat{3}} + \partial_2 \Delta\zeta^3 .$$

And then, we can obtain a metric

$$M_{ab} = \frac{\partial \widetilde{f}^{(a)}}{\partial \varepsilon^{(b)}},$$

where $\varepsilon^{(b)}$ are given by (45); the elements of the metric $M_{ab}$ are

$$M_{11} = M_{12} = M_{13} = M_{14} = M_{15} = M_{16} = 0, M_{17} = \partial_0,$$
$$M_{18} = e_1^{\hat{0}}\partial_0, M_{19} = e_2^{\hat{0}}\partial_0, M_{1,10} = e_3^{\hat{0}}\partial_0;$$
$$M_{21} = -e_1^{\hat{0}}, M_{22} = M_{23} = M_{24} = M_{25} = M_{26} = M_{27} = 0,$$
$$M_{28} = e_1^{\hat{1}}\partial_1, M_{29} = e_2^{\hat{1}}\partial_1, M_{2,10} = e_3^{\hat{1}}\partial_1;$$
$$M_{31} = 0, M_{32} = -e_2^{\hat{0}}, M_{33} = 0, M_{34} = -e_2^{\hat{1}},$$
$$M_{35} = M_{36} = M_{37} = M_{38} = 0, M_{39} = e_2^{\hat{2}}\partial_2, M_{3,10} = e_3^{\hat{2}}\partial_2;$$
$$M_{41} = M_{42} = 0, M_{43} = -e_3^{\hat{0}}, M_{44} = 0, M_{45} = -e_3^{\hat{1}},$$
$$M_{46} = -e_3^{\hat{2}}, M_{47} = M_{48} = M_{49} = 0, M_{4,10} = e_3^{\hat{3}}\partial_3;$$
$$M_{51} = -1, M_{52} = M_{53} = M_{54} = M_{55} = M_{56} = M_{57} = 0,$$
$$M_{58} = e_1^{\hat{1}}\partial_0, M_{59} = e_2^{\hat{1}}\partial_0, M_{5,10} = e_3^{\hat{1}}\partial_0;$$
$$M_{61} = 0, M_{62} = -1, M_{63} = M_{64} = M_{65} = M_{66} = M_{67} = M_{68} = 0,$$
$$M_{69} = e_2^{\hat{2}}\partial_0, M_{6,10} = e_3^{\hat{2}}\partial_0;$$
$$M_{71} = M_{72} = 0, M_{73} = -1, M_{74} = M_{75} = M_{76} = M_{77} = M_{78} = M_{79} = 0,$$
$$M_{7,10} = e_3^{\hat{3}}\partial_0;$$
$$M_{81} = 0, M_{82} = -e_1^{\hat{0}}, M_{83} = 0, M_{84} = -1, M_{85} = M_{86} = M_{87} = M_{88} = 0,$$
$$M_{89} = \partial_1, M_{8,10} = e_3^{\hat{2}}\partial_1;$$
$$M_{91} = M_{92} = 0, M_{93} = -e_1^{\hat{0}}, M_{94} = 0, M_{95} = -1,$$
$$M_{96} = M_{97} = M_{98} = M_{99} = 0, M_{9,10} = \partial_1;$$
$$M_{10,1} = M_{10,2} = 0, M_{10,3} = -e_2^{\hat{0}}, M_{10,4} = 0, M_{10,5} = -e_2^{\hat{1}}, M_{10,6} = -1,$$
$$M_{10,7} = M_{10,8} = M_{10,9} = 0, M_{10,10} = \partial_2.$$

(47)

The Faddeev-Popov determinant is

$$\det M_{ab} \approx \int \prod_{l,m=1}^{10} \mathrm{D}\overline{C}^m \mathrm{D}C^l \exp\left(\frac{\mathrm{i}}{\hbar c}\int \mathrm{d}^4 x L_{\text{ghost-den}}\right),$$

$$L_{\text{ghost-den}} = \sum_{a,b=1}^{10} \overline{C}^a(x) M_{ab} C^l(x).$$

(48)

Now we can write out immediately the vacuum-vacuum transition amplitude of the pure gravitational field in terms of the Feynman's method of path integral with the ten constraint conditions given by (46), the corresponding Faddeev-Popov determinant and the terms of external sources:

$$Z_{\mathrm{G}} = \frac{1}{N} \int \prod_{\alpha,\mu=0}^{3} \mathrm{D}e_{\mu}^{\hat{\alpha}} \prod_{l,m=1}^{10} \mathrm{D}\overline{C}^{m} \mathrm{D}C^{l} \prod_{a=1}^{10} \delta(f^{(a)}) \exp\left(\frac{\mathrm{i}}{\hbar c} \int \mathrm{d}^{4}x \overline{L}_{\mathrm{G-q-den}}\right),$$

$$\overline{L}_{\mathrm{G-q-den}} = \frac{c^{4}}{8\pi G} \det(e_{\mu}^{\hat{\alpha}}) L_{\mathrm{G-den}} + L_{\mathrm{ghost-den}} + L_{JK-\mathrm{den}},$$
(49)

where $f^{(a)}$, $L_{\mathrm{G-den}}$ and $L_{\mathrm{ghost-den}}$ are given by (46), (11) and (48), respectively; the Lagrange density of the terms of external sources

$$L_{JK-\mathrm{den}} = J_{\hat{\alpha}}^{\mu} e_{\mu}^{\hat{\alpha}} + K^{m} \overline{C}^{m} + \overline{K}^{l} C^{l}. \tag{50}$$

Completing the integrals of $\prod_{a=1}^{10} \delta(f^{(a)})$ in (49), we obtain

$$Z_{\mathrm{G}} = \frac{1}{N} \int \prod_{\alpha,\mu=0}^{3} \mathrm{D}e_{\mu}^{\hat{\alpha}} \prod_{l,m=1}^{10} \mathrm{D}\overline{C}^{m} \mathrm{D}C^{l} \exp\left(\frac{\mathrm{i}}{\hbar c} \int \mathrm{d}^{4}x \overline{L}_{\mathrm{G-q-den-eff}}\right),$$

$$\overline{L}_{\mathrm{G-q-den-eff}} = \overline{L}_{\mathrm{G-q-denf}} \big|_{f^{(a)}=0} = \left(\frac{c^{4}}{8\pi G} \det(e_{\mu}^{\hat{\alpha}}) L_{\mathrm{G-den}} + L_{\mathrm{ghost-den}} + L_{JK-\mathrm{den}}\right)\bigg|_{f^{(a)}=0}.$$
(51)

As well known, the whole dynamics of pure quantized gravitational field are determined completely by (49) or (51).

For gravitational field, we have to add *the theory on measure*, e.g., for infinitesimal coordinate distance $\mathrm{d}x^{\mu}$ at spacetime point $x_{\mathrm{P}}^{\mu}$, the actual measured results of observer is $\mathrm{d}\xi^{\hat{\alpha}} = e_{\mu}^{\hat{\alpha}}(x_{\mathrm{P}}^{\lambda}) \mathrm{d}x^{\mu}$, etc (see §4.5). However, $h_{\mu}^{\hat{\alpha}}(x^{\lambda})$ in $e_{\mu}^{\hat{\alpha}}(x^{\lambda})$ is now quantized, this will bring important influence for the results of observation. On the other hand, just due to the quantized $h_{\mu}^{\hat{\alpha}}(x^{\lambda})$ in $e_{\mu}^{\hat{\alpha}}(x^{\lambda})$ and its influence for the process of observation, we can say that the spacetime is quantized.

## 6 Quantum theory of the interaction of gravitational, electromagnetic and spinor field

For electromagnetic field in (31) and (32), simply choosing the temporal gauge

$$A_{0} = 0, \tag{52}$$

As well known, the temporal gauge is existent and for which there is no the corresponding Faddeev-Popov ghost, now we can write out immediately the vacuum-vacuum transition amplitude of the TGESF

$$Z_{\mathrm{GES}} = \frac{1}{N} \int \prod_{\alpha,\mu=0}^{3} \mathrm{D}e_{\mu}^{\hat{\alpha}} \prod_{l,m=1}^{10} \mathrm{D}\overline{C}^{m} \mathrm{D}C^{l} \prod_{\nu=0}^{3} \mathrm{D}A_{\nu} \mathrm{D}\overline{\Psi} \mathrm{D}\Psi$$

$$\times \prod_{a=1}^{10} \delta(f^{(a)}) \delta(A_{0}) \exp\left(\frac{\mathrm{i}}{\hbar c} \int \mathrm{d}^{4}x \overline{L}_{\mathrm{GES-q-den}}\right),$$

$$\overline{L}_{\mathrm{GES-q-den}} = \overline{L}_{\mathrm{G-q-den}} + \det(e_{\mu}^{\hat{\alpha}}) L_{\mathrm{EM-den}} + \det(e_{\mu}^{\hat{\alpha}}) L_{\mathrm{D-den}} + L_{jk-\mathrm{den}},$$
(53)

where $\overline{L}_{\mathrm{G-q-den}}$, $L_{\mathrm{EM-den}}$ and $L_{\mathrm{D-den}}$ are given by (49), (32) and (30), respectively; the

Lagrange density of the terms of external sources

$$L_{jk-\text{den}} = j^{\mu} A_{\mu} + k \overline{\Psi} + \overline{k} \Psi . \tag{54}$$

Completing the integrals of $\prod_{a=1}^{10} \delta(f^{(a)}) \delta(A_0)$ in (53), we obtain

$$Z_{\text{GES}} = \frac{1}{N} \int \prod_{\alpha,\mu=0}^{3} \mathrm{D} e_{\mu}^{\hat{\alpha}} \prod_{l,m=1}^{10} \mathrm{D} \overline{C}^{m} \mathrm{D} C^{l} \prod_{\nu=0}^{3} \mathrm{D} A_{\nu} \mathrm{D} \overline{\Psi} \mathrm{D} \Psi \exp\left(\frac{\mathrm{i}}{\hbar c} \int \mathrm{d}^4 x \overline{L}_{\text{GES-q-den-eff}}\right),$$

$$\overline{L}_{\text{GES-q-den-eff}} = \overline{L}_{\text{GES-q-den}}\big|_{f^{(a)}=0, A_0=0} \tag{55}$$

$$= \overline{L}_{\text{G-q-den-eff}} + \left(\det(e_{\mu}^{\hat{\alpha}}) L_{\text{EM-den}} + \det(e_{\mu}^{\hat{\alpha}}) L_{\text{D-den}} + L_{jk-\text{den}}\right)\big|_{f^{(a)}=0, A_0=0},$$

where $\overline{L}_{\text{G-q-den-eff}}$ is given by (51).

As well known, the whole dynamics of quantized TGESF is determined completely by (53) or (55).

Thus, the both quantum theory of pure gravitational field and that of TGESF are presented completely. The characters of the theories will be studied in further, for instance, could the theories overcome divergencies? Or could the all divergencies in the theories be absorbed in a field renormalization?

## Appendix  The constraint conditions for $h_{\mu}^{\hat{\alpha}}(x^{\lambda})$

As first, we present some constraint conditions which are more than ten, and then, we discuss the characters of the ten constraint conditions in §4.4.

### A-1  Some constraint conditions

What constraint conditions we present here are based on physical investigation.

① The physical constraint conditions

For the request that the characters of gravitational field at every spacetime point should can be observed by observer, the necessary condition is that observer should can be resting at every spacetime point. As well known, this necessary condition leads to

$$g_{(\alpha)} < 0, \ \alpha = 0, 1, 2, 3;$$

where $g_{(\alpha)}$ are the four principal minors of the metric tensor $g_{\mu\nu}$:

$$g_{(0)} = g_{00}, \ g_{(1)} = \begin{vmatrix} g_{00} & g_{01} \\ g_{10} & g_{11} \end{vmatrix}, \ g_{(2)} = \begin{vmatrix} g_{00} & g_{01} & g_{02} \\ g_{10} & g_{11} & g_{12} \\ g_{20} & g_{21} & g_{22} \end{vmatrix}, \ g_{(3)} \equiv g = \begin{vmatrix} g_{00} & g_{01} & g_{02} & g_{03} \\ g_{10} & g_{11} & g_{12} & g_{13} \\ g_{20} & g_{21} & g_{22} & g_{23} \\ g_{30} & g_{31} & g_{32} & g_{33} \end{vmatrix} . \tag{A1}$$

We therefore choose

$$g_{(\alpha)} + 1 = 0 \tag{A2}$$

as the four of ten constraint conditions. Another choice is

$$g_{(\alpha)} + C_{(\alpha)}^2(x) + \varepsilon_{(\alpha)}^+ = 0, \tag{A3}$$

where $C_{(\alpha)}(x)$, $\alpha = 0,1,2,3$ are four arbitrary *c*-number functions being irrelevant to $g_{\mu\nu}$ or $e_\mu^{\hat{\alpha}}$; for avoiding the case of $g_{(\alpha)}=0$ when $C_{(\alpha)}(x)=0$, we add four arbitrary infinitesimal positive *c*-numbers $\varepsilon_{(\alpha)}^+$, $\alpha = 0,1,2,3$.

It is obvious that the four constraint conditions given by (A2) or (A3) act on $g_{\mu\nu}$, substituting (1) to (A2) or (A3), the four constraint conditions thus act on $e_\mu^{\hat{\alpha}}$.

Is there a general coordinate transformation that ensures that (A2) or (A3) holds for an arbitrary manifold coordinate system? I cannot prove this conclusion. The reason that the four constraint conditions given by (A2) or (A3) are introduced is based completely on taking account of the physical process of observation; we therefore call them *the physical constraint conditions*.

However, it is due to the physical reason that (A2) or (A3) is introduced, maybe we could father (A2) or (A3) upon an arbitrary coordinate system, i.e., for an arbitrary coordinate system, we ask that there exists at least a coordinate transformation making that (A2) or (A3) holds in the new coordinate system. Contrarily, for given metric tensor $g_{\mu\nu}$ in a coordinate system, if there is not any coordinate transformation making that (A2) or (A3) holds in the new coordinate system, then such $g_{\mu\nu}$ has not any physical meaning, because the characters of gravitational field at every spacetime point cannot be observed by observer in principle.

② The resting constraint conditions

Now that observer can be resting at every spacetime point, according to the physical meaning of the tetrad $e_\mu^{\hat{\alpha}}$, for infinitesimal coordinate distance $dx^\mu$ at spacetime point $x_P^\mu$, the actual measured results of observer is $d\xi^{\hat{\alpha}} = e_\mu^{\hat{\alpha}}(x_P^\lambda)dx^\mu$, especially, the spacial part

$$d\xi^{\hat{k}} = e_\mu^{\hat{k}}(x_P^\lambda)dx^\mu = e_0^{\hat{k}}(x_P^\lambda)dx^0 + e_l^{\hat{k}}(x_P^\lambda)dx^l ;$$

and, further, for the spacial fixed point, $dx^l = 0$, if observer is indeed resting at the spacial fixed point, then observer is sure to obtain $d\xi^{\hat{k}} = 0$. This investigation leads to

$$e_0^{\hat{k}} = 0, \; k = 1,2,3. \tag{A4}$$

We call (A4) *the resting constraint conditions*.

It can be can proved that we can always enable (A4) to hold by a local Lorentz transformations for arbitrary $e_\mu^{\hat{\alpha}}$; and, further, notice that there are three equations in (A4), for enabling (A4) to hold, the three of the six independent components in a Lorentz transformation are used only.

The form of (A4) is simple, but it leads to a series of conclusions.

Setting $\mu = \nu = 0$ in (1), considering (A4) we obtain $e_0^{\hat{0}} = \sqrt{-g_{00}}$; and, further, setting $\mu = 0, \nu = k$ in (1), considering (A4) and $e_0^{\hat{0}} = \sqrt{-g_{00}}$ we obtain $e_k^{\hat{0}} = -\dfrac{g_{0k}}{\sqrt{-g_{00}}}$, i.e.,

$$e_\mu^{\hat{0}} = -\frac{g_{0\mu}}{\sqrt{-g_{00}}} \ , \ e_{\hat{0}}^\mu = \frac{1}{\sqrt{-g_{00}}} \delta_0^\mu, \qquad (A5)$$

where (2) is used. We see that $e_\mu^{\hat{0}}$, being one of the four $e_\mu^{\hat{\alpha}}$, is completely determined under the conditions (A4).

Using (A4) and (A5), according to the definition (10) of $F_{\mu\nu\lambda}$ we have

$$F_{0\nu\lambda} = -\frac{1}{2}\frac{g_{0\nu}}{g_{00}}g_{00,\lambda} + \frac{1}{2}(g_{\nu\lambda,0} + g_{0\nu,\lambda} - g_{0\lambda,\nu}),$$

$$F_{k\nu\lambda} = -\frac{1}{2}\frac{g_{0k}g_{0\nu}}{g_{00}^2}g_{00,\lambda} + \frac{g_{0k}}{g_{00}}g_{0\nu,\lambda} - \frac{1}{2}(g_{k\nu,\lambda} + g_{k\lambda,\nu} - g_{\nu\lambda,k}) + \eta_{\hat{a}\hat{b}}e_k^{\hat{a}}e_{\nu,\lambda}^{\hat{b}}. \qquad (A6)$$

We see that $F_{0\nu\lambda}$ can be expressed completely by the metric tensor $g_{\mu\nu}$ in this case.

③ The triangular constraint conditions

We can generalize (A4) to the form

$$e_l^{\hat{k}} = 0, \ k > l; k, l = 1, 2, 3.$$

namely,

$$e_1^{\hat{2}} = 0, e_1^{\hat{3}} = 0, e_2^{\hat{3}} = 0. \qquad (A7)$$

It can be can proved that we can always enable (A7) as well as (A4) at the same time to hold by a local Lorentz transformations for arbitrary $e_\mu^{\hat{\alpha}}$; the six constraint conditions (A4) and (A7) make that all six independent components in a Lorentz transformation are used over.

According to (A7) and using the method obtaining (A5) from (A4), (1) and (2), the all $e_\mu^{\hat{k}}$ and $e_{\hat{k}}^\mu$ can be determined completely by the metric tensor $g_{\mu\nu}$:

$$e_\mu^{\hat{1}} = \frac{1}{\sqrt{g_{(0)}g_{(1)}}}\begin{vmatrix} g_{00} & g_{01} \\ g_{\mu 0} & g_{\mu 1} \end{vmatrix}, \ e_\mu^{\hat{2}} = \frac{1}{\sqrt{g_{(1)}g_{(2)}}}\begin{vmatrix} g_{00} & g_{01} & g_{02} \\ g_{10} & g_{11} & g_{12} \\ g_{\mu 0} & g_{\mu 1} & g_{\mu 2} \end{vmatrix},$$

$$e_\mu^{\hat{3}} = \frac{1}{\sqrt{g_{(2)}g_{(3)}}}\begin{vmatrix} g_{00} & g_{01} & g_{02} & g_{03} \\ g_{10} & g_{11} & g_{12} & g_{13} \\ g_{20} & g_{21} & g_{22} & g_{23} \\ g_{\mu 0} & g_{\mu 1} & g_{\mu 2} & g_{\mu 3} \end{vmatrix} = -\sqrt{\frac{g_{(3)}}{g_{(2)}}}\delta_\mu^3. \qquad (A8)$$

$$e_{\hat{1}}^{\mu} = \frac{-1}{\sqrt{g_{(0)}g_{(1)}}}(\delta_0^{\mu}g_{01} - \delta_1^{\mu}g_{00});$$

$$e_{\hat{2}}^{\mu} = \frac{1}{\sqrt{g_{(1)}g_{(2)}}}\left(\delta_0^{\mu}\begin{vmatrix}g_{10} & g_{11}\\ g_{20} & g_{21}\end{vmatrix} - \delta_1^{\mu}\begin{vmatrix}g_{00} & g_{01}\\ g_{20} & g_{21}\end{vmatrix} + \delta_2^{\mu}\begin{vmatrix}g_{00} & g_{01}\\ g_{10} & g_{11}\end{vmatrix}\right),$$

$$e_{\hat{3}}^{\mu} = \frac{-1}{\sqrt{g_{(2)}g_{(3)}}}\left(\delta_0^{\mu}\begin{vmatrix}g_{00} & g_{01} & g_{02}\\ g_{11} & g_{12} & g_{13}\\ g_{21} & g_{22} & g_{23}\end{vmatrix} - \delta_1^{\mu}\begin{vmatrix}g_{00} & g_{02} & g_{03}\\ g_{10} & g_{12} & g_{13}\\ g_{20} & g_{22} & g_{23}\end{vmatrix}\right. \quad\quad (A9)$$

$$\left. + \delta_2^{\mu}\begin{vmatrix}g_{00} & g_{01} & g_{03}\\ g_{10} & g_{11} & g_{13}\\ g_{20} & g_{21} & g_{23}\end{vmatrix} - \delta_3^{\mu}\begin{vmatrix}g_{00} & g_{01} & g_{02}\\ g_{10} & g_{11} & g_{12}\\ g_{20} & g_{21} & g_{22}\end{vmatrix}\right)$$

$$= \sqrt{\frac{g_{(3)}}{g_{(2)}}}g^{\mu 3}.$$

We call (A7) *the triangular constraint conditions* since the both $e_l^{\hat{k}}$ and $e_{\hat{k}}^l$ make up of a 3×3 triangular matrix, respectively. Combining (A5) with (A8) and (A9) we see that the both $e_{\mu}^{\hat{\alpha}}$ and $e_{\hat{\alpha}}^{\mu}$ make up of a 4×4 triangular matrix under the resting and the triangular constraint conditions.

④ The simplest constraint conditions

Notice that under the resting constraint conditions (A5), $e_{\mu}^{\hat{0}}$ that is one of the four $e_{\mu}^{\hat{\alpha}}$ is completely determined. On the other hand, Ref. [11] points out that, in this case (i.e., one of the four $e_{\mu}^{\hat{\alpha}}$ is completely determined), we can choose other $e_{\mu}^{\hat{k}}, k=1,2,3$ to make the Ricci's coefficients of rotation $r_{\hat{\alpha}\hat{\beta}\hat{\gamma}}$ satisfy

$$r_{\hat{0}\hat{a}\hat{b}} + r_{\hat{0}\hat{b}\hat{a}} = 0, \; a \neq b;\; a,b = 1,2,3. \quad\quad (A10)$$

namely

$$r_{\hat{0}\hat{1}\hat{2}} + r_{\hat{0}\hat{2}\hat{1}} = 0,\; r_{\hat{0}\hat{1}\hat{3}} + r_{\hat{0}\hat{3}\hat{1}} = 0,\; r_{\hat{0}\hat{2}\hat{3}} + r_{\hat{0}\hat{3}\hat{2}} = 0. \quad\quad (A11)$$

In Ref. [11], this is called *the simplest case*, because (A11) makes the independent components of the Ricci's coefficients of rotation to reduce to 21 from at most 24 (see (12)) in the 4-dimensional Riemannian geometry, we therefore call (A11) *the simplest constraint conditions*, although the form of (53) is most complex in all presented constraint conditions.

From the definition (8) of the Ricci's coefficients of rotation, (2) and (A5) we have

$$r_{\hat{0}\hat{a}\hat{b}} = e^{\mu}_{\hat{0},\nu} e_{\hat{a}\mu} e^{\nu}_{\hat{b}} + \Gamma^{\mu}_{\lambda\nu} e^{\lambda}_{\hat{0}} e_{\hat{a}\mu} e^{\nu}_{\hat{b}} = \frac{1}{2} \frac{1}{\sqrt{-g_{00}}} (g_{\mu\nu,0} + g_{0\mu,\nu} - g_{0\nu,\mu}) e^{\mu}_{\hat{a}} e^{\nu}_{\hat{b}},$$

$$\begin{aligned} r_{\hat{0}\hat{a}\hat{b}} + r_{\hat{0}\hat{b}\hat{a}} &= \frac{1}{\sqrt{-g_{00}}} g_{\mu\nu,0} e^{\mu}_{\hat{a}} e^{\nu}_{\hat{b}} = \frac{1}{\sqrt{-g_{00}}} (e^{\hat{\alpha}}_{\mu} e_{\hat{\alpha}\nu})_{,0} e^{\mu}_{\hat{a}} e^{\nu}_{\hat{b}} \\ &= \frac{1}{\sqrt{-g_{00}}} (e_{\hat{a}\lambda,0} e^{\lambda}_{\hat{b}} + e_{\hat{b}\lambda,0} e^{\lambda}_{\hat{a}}) = -\frac{1}{\sqrt{-g_{00}}} (e^{\lambda}_{\hat{a},0} e_{\hat{b}\lambda} + e^{\lambda}_{\hat{b},0} e_{\hat{a}\lambda}) \\ &= -\frac{1}{\sqrt{-g_{00}}} g^{\rho\sigma}{}_{,0} e_{\hat{a}\rho} e_{\hat{b}\sigma} = -\frac{1}{\sqrt{-g_{00}}} g^{lm}{}_{,0} e_{\hat{a}l} e_{\hat{b}m}. \end{aligned} \quad (A12)$$

From (A12) we see that (A11) leads to

$$e^{\lambda}_{\hat{1},0} e_{\hat{2}\lambda} + e^{\lambda}_{\hat{2},0} e_{\hat{1}\lambda} = 0, \quad e^{\lambda}_{\hat{1},0} e_{\hat{3}\lambda} + e^{\lambda}_{\hat{3},0} e_{\hat{1}\lambda} = 0, \quad e^{\lambda}_{\hat{2},0} e_{\hat{3}\lambda} + e^{\lambda}_{\hat{3},0} e_{\hat{2}\lambda} = 0, \quad (A13)$$

where $e_{\hat{a}\lambda} = \eta_{\hat{a}\hat{\beta}} e^{\hat{\beta}}_{\lambda}$, $e^{\lambda}_{\hat{a}} = e^{\lambda}_{\hat{a}}(e^{\hat{\beta}}_{\mu})$ are determined by the equations $e^{\hat{\alpha}}_{\mu} e^{\nu}_{\hat{\alpha}} = \delta^{\nu}_{\mu}$. We see that differing from the algebraic constraint conditions (A2), (A3), (A4) and (A7), (A13) is a differential constraint conditions.

From (A10) and the last expression in (A12) we obtain three independent equations

$$g^{lm}{}_{,0} e^{\hat{a}}_{l} e^{\hat{b}}_{m} = 0, \quad a \neq b; \quad a, b = 1, 2, 3. \quad (A14)$$

From (1) we obtain six independent equations

$$\sum_{a=1}^{3} e^{\hat{a}}_{l} e^{\hat{a}}_{m} = g_{lm} - \eta_{\hat{0}\hat{0}} e^{\hat{0}}_{l} e^{\hat{0}}_{m} = g_{lm} - \frac{g_{0l} g_{0m}}{g_{00}}. \quad (A15)$$

On the other hand, there are the nine unknowns $e^{\hat{a}}_{l}$ $(a, l = 1, 2, 3)$, and they can be determined completely by the equations (A14) and (A15).

### A-2 The choice of a group of constraint conditions in terms of the conditions of existence, completeness and consistency

Now we choose ten constraint conditions from (A2), (A3), (A4), (A7) and (A13), which must satisfy *existence*, *completeness* and *consistency* defined in §4.4.

Either (A4) with (A7) or (A4) with (A13) can reduce to 10 from 16 independent components of $e^{\hat{\alpha}}_{\mu}$, and then, using (A2) or (A3) can reduce further to 6. So, arbitrary one of the following four groups of constraint conditions has completeness

(A4) + (A7) + (A2) ; (A4) + (A7) + (A3) ; (A4) + (A13) + (A2) ; (A4) + (A13) + (A3) .

From the expressions of (A5), (A8) and (A9) we see, if (A2) or (A3) holds, then all (A5), (A8) and (A9) are hold, because (A2) or (A3) makes the every square root in (A5), (A8) and (A9) is significative. So the both groups of constraint conditions (A4) + (A7) + (A2) and (A4) + (A7) + (A3) have the consistency.

We have discussed the existence of (A4) and (A7), and pointed out that it cannot be proved whether there is a general coordinate transformation that ensures that (A2) or (A3) holds for an arbitrary manifold coordinate system. However, if the conclude "(A2) or (A3) holds for an arbitrary manifold coordinate system" is acceptable based on the foregoing physical reason, then the existence of the two groups of constraint conditions (A4) + (A7) + (A2) and (A4) + (A7) + (A3) are obvious.

**A-3 The characters of the ten constraint conditions in §4.4**

We therefore try to choose (A2)+(A4)+(A7) or (A3)+(A4)+(A7) as ten constraint conditions.

However, if (A4) and at the same time (A7) hold, then according to (A1) we obtain

$$g_{(0)} = -(e_0^{\hat{0}})^2, \quad g_{(1)} = -(e_0^{\hat{0}})^2(e_1^{\hat{1}})^2, \quad g_{(2)} = -(e_0^{\hat{0}})^2(e_1^{\hat{1}})^2(e_2^{\hat{2}})^2, \quad g_{(3)} = -(e_0^{\hat{0}})^2(e_1^{\hat{1}})^2(e_2^{\hat{2}})^2(e_3^{\hat{3}})^2,$$

in this case we see that (A2) can be simplified to the form

$$e_\alpha^{\hat{\alpha}} = 1 \quad (\alpha = 0, 1, 2, 3). \tag{A16}$$

(A16), (A4) and (A7) are just (46). This group of constraint conditions has the following characters:

① Notice that there is no any derivative term of $g_{\mu\nu}$ in all (A5), (A8) and (A9), if we use (A8) to substitute $e_\nu^{\hat{k}}$ in the expression of $F_{k\nu\lambda}$ given by (A6), we see that there is no any second derivative term of $g_{\mu\nu}$ in the expression of $F_{\mu\nu\lambda}$; And, further, from the discussion in the Section 2 we know that $F_{\mu\nu\lambda}$ is a tensor in the global manifold coordinate system when the local frame coordinate system is fixed, that means that if we start from (A6) in which $e_\nu^{\hat{k}}$ is substituted by (A8), after an arbitrary general coordinate transformation, what the expression of $F_{\mu\nu\lambda}$ we obtain is still only a function of $g_{\mu\nu}$ and the first derivative of $g_{\mu\nu}$. That means that if (A4) and at the same time (A7) hold, then $F_{\mu\nu\lambda}$ is only a function of $g_{\mu\nu}$ and the first derivative of $g_{\mu\nu}$:

$$F_{\mu\nu\lambda} = F_{\mu\nu\lambda}(g_{\rho\sigma}, \partial_\zeta g_{\rho\sigma}).$$

② It is obvious that (A16) does not depend upon prescribing the temporal axis in the 4-dimensional spacetime. On the other hand, although it is due to the physical reason that (A4) is introduced, the elements of $e_\mu^{\hat{\alpha}}$ at last make up of a 4×4 triangular matrix under the constraint conditions (A4) and (A7). On the other hand, according to an algebraical theorem, an arbitrary symmetrical $n \times n$ matrix $g_{\mu\nu}$ can always be written in the product of a triangular matrix and its transposed matrix if and only if the all principal minors $g_{(\alpha)}$ ($\alpha = 0, 1, 2, \cdots, n-1$) of $g_{\mu\nu}$ do not equalize zero. (In fact, (1), (A5), (A8) and (A9) can be seemed as a proof of the theorem when $n = 4$.) Thus, we can choose other conditions substituting (A4) and (A7) to make the elements of $e_\mu^{\hat{\alpha}}$ make up of a 4×4 triangular matrix, that means that (A4) and (A7) do not depend upon prescribing the temporal axis in the 4-dimensional spacetime in principle. So the ten constraint conditions (46) do not depend upon prescribing the temporal axis in the 4-dimensional spacetime.